\RequirePackage{fix-cm}
\documentclass[fleqn,usenatbib]{mnras}
\usepackage[dvipsnames]{xcolor}
\usepackage{eso-pic}

\AddToShipoutPictureBG*{%
  \AtPageUpperLeft{%
    \hspace{0.75\paperwidth}%
    \raisebox{-1.\baselineskip}{%
      \makebox[0pt][l]{\textnormal{DES-2025-0961}}
}}}%

\AddToShipoutPictureBG*{%
  \AtPageUpperLeft{%
    \hspace{0.75\paperwidth}%
    \raisebox{-2.\baselineskip}{%
      \makebox[0pt][l]{\textnormal{FERMILAB-PUB-26-0075-V}}
}}}%

\usepackage{newtxtext,newtxmath}
\usepackage{mdframed}
\usepackage{graphicx}
\usepackage{subcaption}
\usepackage{dsfont}
\usepackage{lineno}
\let\oldequation\equation
\let\oldendequation\endequation
\renewenvironment{equation}
  {\linenomathNonumbers\oldequation}
  {\oldendequation\endlinenomath}

\usepackage{hyperref}
\usepackage[T1]{fontenc}
\usepackage{ae,aecompl}
\usepackage{todonotes}
\usepackage{enumitem}

\usepackage{amsmath}	
\usepackage{tikz-cd}
\usepackage{mathtools}
\usepackage{amsfonts}
\usepackage{tabularx}
\usepackage{algorithmicx}
\usepackage{algpseudocode}
\usepackage{algorithm}
\usepackage{natbib}
\usepackage{booktabs}

\newcommand{\healpix}{\textsc{HEALPix}}
\newcommand{\pkdgrav}[0]{\texttt{PKDGRAV3}}
\newcommand{\concept}[0]{\textsc{concept}}

\newcommand{\hyperrank}[0]{\textsc{hyperrank}}
\newcommand{\Om}{\Omega_{\rm m}}

\newcommand{\VICreg}[0]{{VICReg}}
\newcommand{\TARP}[0]{{\texttt{TARP}}}

\newcommand{\mathd}{\ensuremath{\mathrm{d}}} 
\newcommand{\nside}[1]{$\textsc{nside} = #1$}

\newcommand{\cond}[0]{\ensuremath{\, | \,}}
\newcommand{\iintt}[0]{\ensuremath{\int \! \! \! \! \int}}
\newcommand{\simsampdist}[0]{\ensuremath{p^{\textrm{sim}}}}
\newcommand{\tarpprior}[0]{\ensuremath{p^{\textrm{\TARP{}}}}}

\newcommand{\imagunit}{\ensuremath{\textrm{i}\mkern1mu}}
\newcommand{\direction}{\ensuremath{\hat{n}}}

\newcommand{\subsubsect}[1]{\smallskip \noindent \textbf{#1:}}

\newcommand{\dat}{\textbf{x}}
\newcommand{\parvec}{\boldsymbol{\theta}} 
\newcommand{\xvec}{\textbf{{x}}}

\newcommand{\lossfunction}{\mathcal{L}}

\DeclareMathOperator*{\avg}{Avg}
\DeclareMathOperator*{\argmin}{argmin}
\DeclareMathOperator*{\argmax}{argmax}
\DeclareMathOperator*{\MAP}{MAP}
\DeclareMathOperator*{\FOM}{FOM}

\usepackage[normalem]{ulem}

\definecolor{darkgreen}{rgb}{0.0,0.66,0.37}

\usepackage{multirow}

\makeatletter
\renewcommand{\@fnsymbol}[1]{%
  \ifcase#1\or
    *\or
    \dagger\or
    \mathsection\or
    \ddagger\or
    \mathparagraph\or
    \ | \or
    **\or
    \dagger\dagger\or
    \ddagger\ddagger
  \fi
}
\makeatother

\newcounter{sharefoot}

\makeatletter
\newcommand{\titlefnsuperscript}{%
  \renewcommand\@makefnmark{\hbox{\@textsuperscript{\normalfont\@thefnmark}}}%
}

\newcommand{\titlefnbaseline}{%
  \renewcommand\@makefnmark{\hbox{\normalfont\@thefnmark}}%
}

\newcommand{\baselineThanks}[1]{%
  \begingroup\titlefnbaseline\thanks{#1}\endgroup
}

\DeclareRobustCommand{\sharefnmark}{%
  \hyperlink{Hfootnote.\number\value{sharefoot}}{%
    \footnotemark[\value{sharefoot}]%
  }%
}
\makeatother

\title[DES Year 3: simulation-based $w$CDM inference]{Dark Energy Survey Year 3 results: optimized $w$CDM simulation-based inference with weak lensing map-level hybrid statistics}

\author[J. Williamson \& T. L. Makinen et al.]{
\parbox{\textwidth}{
\large
J.~Williamson \baselineThanks{E-mail: joshua.williamson.21@ucl.ac.uk}$^{1}$
\ \&\ \ T.~L.~Makinen \baselineThanks{E-mail: tlm41@cam.ac.uk}$^{2,3}$\thanks{DES External Collaborators}
\setcounter{sharefoot}{\value{footnote}}%
,  N.~Porqueres,$^{4}$\sharefnmark \
N.~Jeffrey,$^{1,5}$
A.~Heavens,$^{2}$\sharefnmark \
M.~Gatti,$^{6,7}$
B.~D.~Wandelt,$^{8,9}$\sharefnmark \
L.~Whiteway,$^{1}$
J.~Prat,$^{10,11}$
A.~Alarcon,$^{6}$
A.~Amon,$^{12}$
K.~Bechtol,$^{13}$
M.~R.~Becker,$^{14}$
G.~M.~Bernstein,$^{15}$
A.~Campos,$^{16,17}$
A.~Carnero~Rosell,$^{18,19,20}$
R.~Chen,$^{21}$
A.~Choi,$^{22}$
J.~DeRose,$^{23}$
C.~Doux,$^{15,24}$
A.~Drlica-Wagner,$^{25,26,7}$
K.~Eckert,$^{15}$
S.~Everett,$^{27}$
A.~Fert\'e,$^{28}$
Z.~Gong,$^{29}$
D.~Gruen,$^{29}$
R.~A.~Gruendl,$^{30,31}$
K.~Herner,$^{26}$
M.~Jarvis,$^{15}$
T.~Kacprzak,$^{32}$
O.~Lahav,$^{1}$
J.~McCullough,$^{12,33,28,29}$
J.~Myles,$^{12}$
A.~Navarro-Alsina,$^{34}$
S.~Pandey,$^{15}$
M.~Raveri,$^{35}$
R.~P.~Rollins,$^{36}$
E.~S.~Rykoff,$^{33,28}$
C.~S{\'a}nchez,$^{37,38}$
L.~F.~Secco,$^{7}$
I.~Sevilla-Noarbe,$^{39}$
E.~Sheldon,$^{40}$
T.~Shin,$^{41}$
A.~Thomsen,$^{32}$
M.~A.~Troxel,$^{21}$
I.~Tutusaus,$^{42}$
T.~N.~Varga,$^{43,44,45}$
B.~Yanny,$^{26}$
B.~Yin,$^{21}$
J.~Zuntz,$^{46}$
T.~M.~C.~Abbott,$^{47}$
M.~Aguena,$^{48,19}$
F.~Andrade-Oliveira,$^{49}$
D.~Brooks,$^{1}$
R.~Camilleri,$^{50}$
J.~Carretero,$^{38}$
R.~Cawthon,$^{51}$
M.~Crocce,$^{52,6}$
L.~N.~da Costa,$^{19}$
T.~M.~Davis,$^{50}$
J.~De~Vicente,$^{39}$
S.~Desai,$^{53}$
H.~T.~Diehl,$^{26}$
B.~Flaugher,$^{26}$
J.~Frieman,$^{25,26,7}$
J.~Garc\'ia-Bellido,$^{54}$
G.~Gutierrez,$^{26}$
S.~R.~Hinton,$^{50}$
D.~L.~Hollowood,$^{55}$
K.~Kuehn,$^{56,57}$
J.~L.~Marshall,$^{58}$
J.~Mena-Fern{\'a}ndez,$^{59,24}$
R.~Miquel,$^{60,38}$
J.~J.~Mohr,$^{29}$
J.~Muir,$^{61}$
A.~A.~Plazas~Malag\'on,$^{33,28}$
A.~Porredon,$^{39,62}$
A.~Roodman,$^{33,28}$
E.~Sanchez,$^{39}$
D.~Sanchez Cid,$^{39,49}$
E.~Suchyta,$^{63}$
M.~E.~C.~Swanson,$^{30}$
C.~To,$^{25}$
D.~L.~Tucker,$^{26}$
V.~Vikram,$^{14}$
A.~R.~Walker,$^{47}$
N.~Weaverdyck,$^{64,23}$
J.~Weller$^{44,45}$
\begin{center}(DES Collaboration)\end{center}
}\\
\parbox{\textwidth}{\small
\textit{The authors' affiliations are shown in Appendix~\ref{append:affiliations}. \vspace{-0.5cm}}
}}

\date{Accepted XXX. Received 2025; in original form 2025}
\pubyear{2025}

\begin{document}
\label{firstpage}
\pagerange{\pageref{firstpage}--\pageref{lastpage}}

\begingroup
  \titlefnsuperscript
  \maketitle
\endgroup

\begin{abstract}
We present cosmological constraints from the Dark Energy Survey Year 3 (DES Y3) weak lensing data using hierarchical hybrid statistics within a Bayesian simulation-based inference framework that is based on the Gower Street simulations. To maximize the precision of the inference, we have developed a new, information-theory based, data compression of the weak lensing maps to just seven highly informative summary statistics. The hybrid scheme exploits the high information content of the power spectrum, compressing both the power spectrum and neural-based summaries that are designed to extract further information. Our simulation-based approach enables principled forward modelling of all major sources of systematic uncertainty and survey properties into realistic mock observations. The inference is validated through coverage tests and checks for robustness against baryonic feedback. Assuming a $w$CDM cosmology, our analysis yields $S_8 = 0.808 \pm 0.017$, $\Om = 0.325 \pm 0.024$, and $w < -0.766$ (marginalized posterior 68 per cent credible intervals). This rigorous combination of information theory, physics- and neural network-based extreme data compression, and principled Bayesian analysis improves the figure of merit for $(\Om, S_8, w)$ by 60 per cent over the previous state-of-the-art, and by almost a factor of 3 over two-point analyses of the same data. The results are in good agreement with Planck 2018 TTTEEE+lowE CMB inferred values of $\Om, S_8$, and $w$, and are consistent with $\Lambda$CDM cosmology. They are the most precise joint constraints on $(\Om, S_8, w)$ from weak gravitational lensing data alone of any survey to date. 
\end{abstract}



\section{Introduction}

Inhomogeneities in the matter distribution in the Universe distort light paths from distant galaxies, producing per cent-level distortions to galaxy shapes. This distortion pattern, known as cosmic shear, provides valuable cosmological information, with particular sensitivities to the average density of matter and to its clumpiness. It probes the growth rate of structure and the geometry of the Universe, indirectly yielding information on the nature of dark energy.

Stage III photometric surveys \citep{albrecht2006reportdarkenergytask} such as the Dark Energy Survey \citep[DES,][]{DESy6-3x2pt}, the Kilo-Degree Survey \citep[KiDS,][]{KIDS_3x2pt}, and Hyper Suprime-Cam \citep[HSC,][]{HSC_3x2pt} have now delivered per cent-level weak lensing constraints on the amplitude of matter fluctuations $\sigma_8$ and the matter density $\Om$. The cosmological information in these analyses has been extracted through two-point statistics of weak lensing shear fields i.e. using the statistics that, if the underlying field is Gaussian, are optimal. However, non-linear structure formation makes the late-time matter distribution in the Universe highly non-Gaussian, and several studies from Stage III collaborations infer cosmological constraints using higher-order statistics that can capture the non-Gaussian information \citep{Jeffrey_massmap_vmim,gatti_dark_2024,juditprat_dark_2025, jeffrey2024dark, Burger_KiDS_2024, HSC_SBI, zuercherpeaks}. 

Unfortunately, inferring cosmological information using higher-order statistics in a Bayesian analysis is difficult because of a lack of analytical modelling. Deriving a closed-form expression for the sampling distribution of higher-order statistics is either prohibitively computationally expensive or completely intractable. In recent years, advances in cosmological simulations and forward modelling of survey systematics have made possible the computation of these likelihoods through simulation-based inference. By explicitly modelling all statistical assumptions, uncertainties, and systematics (such as masks and noise variations) into mock observables within these simulations, and then compressing these into summaries using neural networks, we can robustly estimate the likelihood function in our cosmological inference.

The challenge of extracting cosmological information beyond the power spectrum has motivated substantial methodological development in data compression for map-level inference. Massively Optimized Parameter Estimation and Data compression (MOPED; \citealt{moped}) demonstrated that high-dimensional data can, under Gaussian assumptions, be compressed to the extreme (i.e. to where the number of summaries equals the number of parameters) while simultaneously preserving Fisher information. This framework was extended to non-Gaussian data through score compression \citep{alsing2018_general}, which projects data onto gradients of the log-likelihood. Neural approaches have built upon these foundations, with Information Maximizing Neural Networks (IMNNs; \citealt{Charnock_IMNN}, \citealt{makinen2021}) training networks to find non-linear compressions that maximize Fisher information about cosmological parameters at a fiducial point, and with Variational Mutual Information Maximization \citep[VMIM,][]{Jeffrey_massmap_vmim, barber2003imalgorithm,poole2019variationalboundsmutualinformation} maximizing the expected information over the prior. A recent comprehensive comparison of neural compression \citep{lanzieri2024optimalneuralsummarisationfullfield} demonstrated VMIM achieves optimal compression by recovering 100 per cent of the full-field information in correlated fields, compared to only 81 per cent for standard regression-based approaches. The VMIM approach was first applied to DES (Science Verification) data in \cite{Jeffrey_massmap_vmim}. Complementary to these learned compressions, interpretable beyond-two-point statistics have proven able to extract additional cosmological information more effectively than two-point only: in DES, \cite{marco_ST_WPH_24} used second- and third-order moments, scattering transforms, and wavelet phase harmonics to constrain $w$CDM cosmology. In parallel, \cite{juditprat_dark_2025} used Betti numbers and persistent homology, while \cite{jeffrey2024dark} used peak counts and convolutional neural networks (CNNs) at the map-level. \cite{HSC_SBI} provided an analysis of HSC data using peaks and minima counts with Minkowski functionals. In KiDS, \cite{Burger_KiDS_2024} also provided constraints using second- and third-order statistics. 

Hybrid summary statistics \citep[for existing work see][]{makinen2024hybridsummarystatistics, Makinen_2025_hybridfisher} build a principled framework for combining these interpretable physics-based summaries with flexible neural network compressions. \cite{Ntampaka_2020} showed that by stacking power spectrum and map-level CNN galaxy clustering observables into the same neural network, there was no precision improvement over using the map-level CNN only. By optimizing neural networks to maximize the conditional mutual information between data and parameters given existing statistics, this approach ensures that map-based features complement rather than duplicate information already captured by traditional two-point functions. This information-theoretic compression scheme has been demonstrated in simulation studies to constrain $\Om$ and $S_8$ significantly more than angular power spectra alone \citep{Makinen_2025_hybridfisher}.

This machine-learning technique is not a `black-box' solution for the cosmological parameters.  The neural compression is used simply to define highly informative statistics (i.e. functions of the data) that are conceptually little different from more traditional compression to power spectra or other summaries.  The distinction is that these statistics are optimized to maximize information content. The unknown sampling distribution is dealt with in a simple Bayesian neural density estimation procedure that is learned from forward modelling mock observations, whereas traditional Bayesian analysis assumes a likelihood function, typically Gaussian, which may not hold in detail.

This paper has one main scientific aim: to demonstrate that we can obtain more precise $w$CDM constraints from weak lensing map-level inference by leveraging a new and optimized information-theoretic data compression scheme: \emph{hierarchical hybrid statistics}. This work directly extends \cite{jeffrey2024dark}, using the same analysis pipeline and implementing hybrid statistics as the optimization objective.

With this powerful new method, optimized to extract maximal cosmological information, we unify interpretable two-point correlations with flexible map-based features in a single simulation-based inference (SBI) framework, yielding robust constraints for DES~Y3 and directly improving the future analysis pathway for next-generation data from surveys such as \textit{Euclid} and the Vera C.\ Rubin Observatory (LSST).

We review simulation-based inference (SBI) and neural density estimation in Section~\ref{sec:sbi}, describing the framework of SBI and neural density estimation. In Section~\ref{sec:method} we introduce and derive the hybrid statistics method. We explain in Section~\ref{sec:weaklensing} the theoretical model of the observable (i.e. weak gravitational lensing). Section~\ref{sec:simsdata} describes the simulations used in this work, the DES Y3 data, and the forward modelling of survey properties into realistic mock weak lensing data. Section~\ref{sec:compression} details the SBI analysis pipeline and compression scheme; Section~\ref{sec:validation} outlines validation tests; Section~\ref{sec:results} presents our DES Y3 cosmological results and a discussion of our analysis. Our conclusions are in Section~\ref{sec:conclusion}.

\section{Simulation-based inference}\label{sec:sbi}
The extraction of cosmological information from weak lensing data relies fundamentally upon our ability to characterize the sampling distribution of chosen summary statistics. Two-point statistics such as the angular power spectrum admit analytic likelihood functions only under the assumption of Gaussian fields; this assumption breaks down for the late-time matter distribution (which is manifestly non-Gaussian due to non-linear gravitational evolution), and breaks down as well when one properly accounts for observational systematics. For higher-order statistics, the sampling distribution is not generally known and is analytically intractable. In this context the classical approach of assuming a Gaussian likelihood with an analytic covariance matrix is an approximation of uncertain validity.

Simulation-based inference (SBI) provides a principled framework for Bayesian inference when the likelihood function $p(\dat \cond \parvec)$ (for parameters $\parvec$ given observed data $\dat$) is intractable, but when forward simulations from the data-generating process are available \citep{Cranmer30055}. Rather than imposing a functional form on the likelihood, SBI methods learn the statistical relationship between parameters and observables directly from simulated data. This approach naturally accommodates the full complexity of the forward model, including all sources of systematic uncertainty, without requiring analytic approximations. In this sense, SBI is a more principled approach to modelling observations than traditional analytic likelihood methods, as it explicitly defines all statistical assumptions through the forward simulation pipeline.

The goal of Bayesian inference is to obtain the posterior distribution of the parameters, which by Bayes' theorem is expressed as
\begin{equation}\label{eq:bayes}
    p(\parvec \cond \dat) = \frac{p(\dat \cond \parvec) \, p(\parvec)}{p(\dat)},
\end{equation}
where $p(\dat \cond \parvec)$ is the likelihood (or sampling distribution), $p(\parvec)$ is the prior, and $p(\dat) = \int p(\dat \cond \parvec) p(\parvec) \, \mathd \parvec$ is the evidence. When the likelihood is intractable, SBI methods circumvent its direct evaluation by instead learning either the likelihood, the posterior, or the likelihood-to-evidence ratio from $N$ simulations of parameter-data pairs of vectors $\{(\parvec_i, \dat_i)\}_{i=1}^{N}$.

For high-dimensional data $\dat$, SBI can be decomposed into three stages: (i) forward modelling of the data-generating process, (ii) compression of high-dimensional data to informative low-dimensional summaries, and (iii) density estimation to characterize the statistical relationship between parameters and summaries. Here we review the formalism for data compression and density estimation, deferring until Section~\ref{sec:method} a discussion of our new hierarchical hybrid compression framework.

\subsection{Data compression}

Estimating the likelihood $p(\dat \cond \parvec)$ becomes increasingly intractable as the dimension of $\dat$ increases. See \cite{Dai_2022, Mudur_2024} and references therein for studies addressing this. In this DES weak lensing analysis, the data dimensionality is $\sim 10^7$ for the case of map-level inference and $10^3$ for inference using standard cosmological statistics such as power spectra. We wish to find some (deterministic) function of the data, $F: \dat \mapsto \textbf{t}$, that reduces dimensionality while preserving information about the parameters of interest $\parvec$. In some cases, it is possible to find $F$ for which the summary dimensionality of $\textbf{t}$ is the same as the number of parameters, and such that the (local) Fisher information is preserved \citep{moped, alsing2018_general}. The objective of this work is to exploit the scientific benefits of a new, optimized neural compression function to better combine existing and map-level statistics for DES Y3 data. We hierarchically optimize neural networks via a mutual information criterion to compress the DES Y3 data products in a complementary and parameter-information maximizing fashion. We present this method in full in Section~\ref{sec:method}.

\subsection{Neural density estimation}

A neural density estimator \citep[NDE; e.g.][]{bishop_mdn_1994} uses a neural network to give an estimate $q$ of a desired (conditional) probability distribution $p$. We discuss two types of NDEs:

\begin{itemize}[wide]
\item{
    Neural Posterior Estimation (NPE; \citealt{papamakarios2018fastepsilonfreeinferencesimulation}) directly parameterizes an approximation $q_\varphi(\parvec \cond \textbf{t})$ to the posterior using conditional density estimators. Sampling from the NPE posterior is direct and inexpensive, in contrast to NLE (described below), but is confined to a single prior choice.
}
\item{
    Neural Likelihood Estimation (NLE; \citealt{papamakarios2019sequential, pydelfiAlsing_2019}) learns a surrogate likelihood $q_\varphi(\textbf{t} \cond \parvec)$, which can subsequently be combined with an arbitrary  prior and sampled via standard MCMC methods. NLE varies weights and biases (parameterized as $\varphi$) to minimize the loss
    \begin{equation}\label{eq:nde_loss}
        U(\boldsymbol{\varphi}) = - \sum_{i=1}^{N} \ln q( \textbf{t}_i \cond \parvec_i ; \boldsymbol{\varphi}),
    \end{equation}
    over a set of $N$ parameter-data simulated pairs $\{(\parvec_i, \dat_i)\}_{i=1}^{N}$. Here the simulation parameters $\parvec_i$ are sampled from a \emph{simulation sampling distribution} $\simsampdist(\parvec)$. Minimizing the loss $U$ is equivalent to minimizing the information gain of the target distribution over and above $q$ (i.e. minimizing the Kullback-Leibler divergence).
}
\end{itemize}

This work uses NLE, as this permits the simulation sampling distribution to differ from the analysis prior (with the analysis prior applied only at the inference stage); we show in Section~\ref{sec:simsdata} that our need for simulation efficiency makes this flexibility particularly advantageous. 
We use Masked Autoregressive Flows \citep[MAF;][]{papa_mafs2017} and Mixture Density Networks \citep[MDN;][]{bishop_mdn_1994} to model a surrogate conditional probability function. For inference, we use an ensemble of estimators for the implicit likelihood (see Subsection~\ref{subsec:nde_details} for details). It is worth noting that in this paper we assign $q$ to mean any learned conditional probability distribution in general, being either NLE or NPE. Both are implemented in this analysis with different MAF and MDN architectures.

\section{Hierarchical hybrid statistics and compression}\label{sec:method}
Our domain knowledge can guide us to define summary statistics that capture much of the information content in a dataset (e.g. the two-point function in weak lensing cosmology). Moreover, neural networks can have difficulty capturing Gaussian field information with limited training data \citep{makinen2021,makinen2024hybridsummarystatistics}. Here we present a method that ensures \textit{complementary} neural information capture in the form of \textit{hybrid summary statistics}. We describe our information-theoretic compression method, formalizing techniques presented in \citet{makinen2024hybridsummarystatistics, Makinen_2025_hybridfisher} and extending hybrid statistics to a hierarchy of data compression tasks.

\subsection{Differential entropy and mutual information}

We recall standard definitions from information theory. The \textit{differential entropy} of a random variable is

\begin{equation}
    h(A) = - \int \!  p(a) \ln(p(a)) \, \mathd a ,
\end{equation}
while the \textit{mutual information} between two random variables is the information gain of the joint distribution over and above the product of the marginals (i.e. is a Kullback-Leibler divergence):
\begin{equation}
    I(A; B) = \iintt p(a, b) \ln \left( \frac{p(a, b)}{p(a)p(b)} \right) \mathd a \, \mathd b .
\end{equation}
Conditioning on a third random variable gives the \textit{conditional entropy}
\begin{equation}
    h(A \cond C) = -\int p(a, c) \ln p(a \cond c) \, \mathd a \, \mathd c
\end{equation}
and the \textit{conditional mutual information}
\begin{multline}
    I(A; B \cond C) = \\
    \int p(c) \iintt p(a, b \cond c) \ln \left( \frac{p(a,b \cond c)}{p(a \cond c)p(b \cond c)} \right)   \mathd a \, \mathd b \, \mathd c \, .
\end{multline}
Straightforward calculations show
\begin{equation}
    I(A; B) = h(A) - h(A \cond B) \label{eq:mi-entropy-uncond}
\end{equation}
and
\begin{equation}
    I(A; B \cond C) = h(A \cond C) - h(A \cond B, C).\label{eq:mi-entropy-cond}
\end{equation}

\subsection{Variational mutual information maximization}

We use the above definitions in a Bayesian inference problem with $A$ being the parameters of interest $\Theta$ and $B$ being the data $\xvec$. If we wish to replace $\xvec$ with a summary $F(\xvec)$ where $F$ is some deterministic compressor function, then which $F$ is best? The VMIM framework \citep{Jeffrey_massmap_vmim} finds a compressor function $F^*$ (implemented as a neural network) that maximizes the mutual information $I(\Theta; F(\xvec))$ between the parameters and the compressed data. In equation (\ref{eq:mi-entropy-uncond}) the first term on the right hand side is independent of $B$, so varying $F$ to maximize mutual information is the same as varying $F$ to minimize differential entropy:
\begin{equation}
    F^* \equiv \argmax_{F} I(\Theta; F(\xvec)) = \argmin_{F} h(\Theta \cond F(\xvec)); \label{eq:f_star}
\end{equation}
there is an equivalent definition/result for the conditional case. Equivalently, we could minimize the \textit{expected posterior entropy} \citep{hoffmann2023minimising}. In what follows, we will consistently use an asterisk to denote this `mutual-information-maximizing' compression function.

We have $I(\Theta; F^*(\xvec)) \le I(\Theta; \xvec)$ because compression cannot increase information (the \textit{data processing inequality}). Equality will hold only in special cases (in which case $F^*$ is said to be \textit{sufficient}); despite this, the optimization in equation (\ref{eq:f_star}) is useful in practice. 

We will find $F^*$ by minimizing the differential entropy; how then to actually calculate $h(\Theta \cond F(\xvec))$? Our simulation suite provides samples $\parvec_i$ and $\xvec_i$ (the forward model adds noise, so the relationship between these two is not deterministic); from $\xvec_i$ we calculate $F(\xvec_i)$. As commented earlier, the simulation sampling distribution $\simsampdist(\parvec)$ need not be identical to the parameter prior used when calculating the posterior; however, the support of the former should enclose the support of the latter. We may use these samples to estimate the differential entropy, and thus to define a loss function $\lossfunction{}$:
\begin{equation}
    h(\Theta \cond F(\xvec)) \approx  - \avg_i \left[ \ln \tilde{q}(\parvec_i \cond F(\xvec_i)) \right] \equiv \lossfunction{}(\tilde{q}, F).
    \label{eq:loss_function}
\end{equation}
Here $\tilde{q}$ is a neural network that approximates the posterior kernel i.e. the conditional density $p(\parvec \cond F(\xvec))$. The parameters of $\tilde{q}$ are optimized (so as to provide a good match to the posterior kernel) simultaneously with the parameters of $F$ being optimized (so as to minimize the differential entropy). We can achieve this double optimization using the single loss function $\lossfunction{}$, which is explicitly a function of both sets of parameters (i.e. for $\tilde{q}$ and for $F$). The dimensionality of each $\textbf{z}_j = F_j(\xvec_j)$ is a hyperparameter that controls the bottleneck of the information flow to the density estimator $q(\parvec \cond \cdot)$. 

The averaging over simulations in the middle term in equation (\ref{eq:loss_function}) is intended to approximate an expectation with respect to the probability density $p(\parvec, \xvec)$. However, the simulations are actually a sample from the somewhat different density $\simsampdist(\parvec, \xvec) \equiv p(\xvec \cond \parvec) \simsampdist(\parvec)$; this increases the error in our approximation of the differential entropy.
For the cosmological problem considered in this paper, the dataset is too large to be efficiently passed to a single conventional neural network on standard devices for training in batches. A practical solution is found by extending hybrid statistics to a hierarchical, \textit{conditional} optimization problem, as described in Subsection~\ref{subsec:hierarchicalcompression}.

\subsection{Hierarchical data compression} \label{subsec:hierarchicalcompression}

Consider the analysis of weak lensing data in which $\Theta$ are cosmological parameters and the data $\xvec$ are field data (e.g. a convergence $\kappa$ map). The power spectrum $C_\ell(\xvec)$ is a useful (i.e. information rich) summary statistic (i.e. compression of the data); however, the distribution $\xvec$ is non-Gaussian so there is additional information. Our hybrid scheme produces an optimized compression of the data derived jointly from physically motivated summaries (i.e. the power spectra) and, conditional on the power spectra, from CNN-based summaries. To avoid computing limitations we partition the survey area into patches and calculate the CNN-based summaries separately on each patch. This partitioning loses large scale information, but this is not a problem as this information is essentially Gaussian and hence is already present in the power spectra.

\begin{figure}
    \centering
    \includegraphics[width=0.6\linewidth]{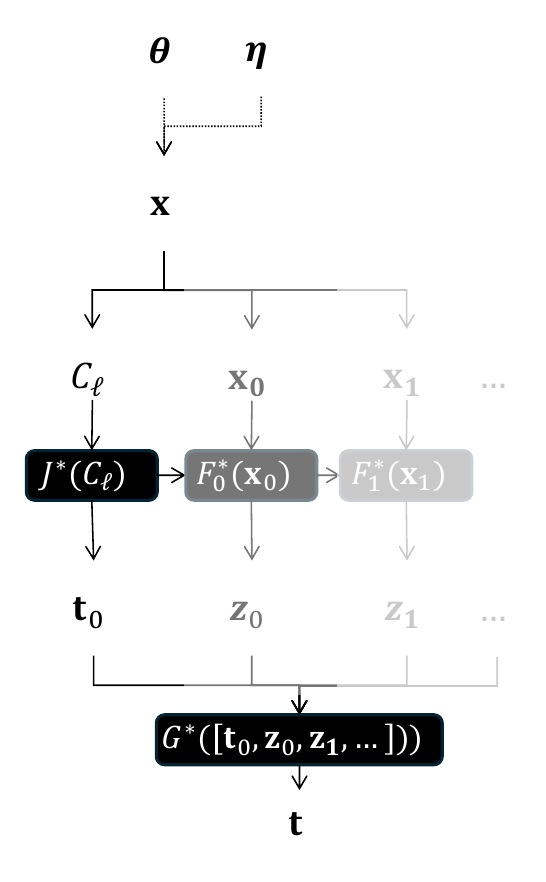}
    \caption{Plate diagram illustrating the hierarchical hybrid compression of data $\xvec$ with parameters $\parvec$ and nuisance parameters $\eta$. The data are partitioned into subsets $\xvec_0, \xvec_1, \dots$, and the power spectrum $C_\ell$ is computed from the full field. Compressions are performed sequentially: $J^*(C_\ell)$ compresses the power spectrum, then each $F^*(\xvec_i)$ compresses a data subset conditional on all previously obtained summaries (indicated by the horizontal arrows). The shading of each compression block reflects the conditioning hierarchy, from darkest (no conditioning) to lightest (conditioned on all prior summaries). A final regularized compression $G^*([\textbf{t}_0, \textbf{z}_0, \dots])$ aggregates all summaries into a low-dimensional statistic $\textbf{t}$.}
    \label{fig:plate3}
\end{figure}

As before, we assume that we have samples $(\parvec_i, \xvec_i)$ from the joint distribution $p(\parvec, \xvec)$. See Fig.~\ref{fig:plate3}. We proceed in several steps emphasizing that a compression denoted with a star represents a network being optimized and a compression function without a star is amortized.

\begin{enumerate}[wide]

    \item{Partition the data vector $\xvec$ into subsets $\xvec_A, \xvec_B, \xvec_C$. Recall that the data vector is a map, and so this step corresponds to partitioning the map into subregions.}
    
    \item{ Use VMIM to find a compression $J^*(C_\ell)$ of the power spectra that maximizes the mutual information $I(\parvec; J^*(C_\ell))$.}
    
    \item{Find a compression $F_A^*(\xvec_A)$ of the first data subset that maximizes the conditional mutual information $I(\parvec ; F_A(\xvec_A) \cond J^*(C_\ell))$.}
    
    \item{Repeat the previous step for the remaining data subsets, each time conditioning on the compressed data obtained so far. We thereby obtain $F_B^*(\xvec_B)$ maximizing the conditional mutual information $I(\parvec ; F_B(\xvec_B) \cond J^*(C_\ell), F_A^*(\xvec_A))$, and $F_C^*(\xvec_C)$ maximizing the conditional mutual information $I(\parvec ; F_C(\xvec_C) \cond J^*(C_\ell), F_A^*(\xvec_A), F_B^*(\xvec_B))$}.
    
    \item{Steps (ii)-(iv) are repeated with different network initializations, and the concatenated summaries undergo a final (unconditional) VMIM compression $G^*$ of all the compressed results so far, obtaining $G^*(\xvec)$ that maximizes the mutual information $I(\parvec ; J^*(C_\ell), F_A^*(\xvec_A), F_B^*(\xvec_B), F_C^*(\xvec_C))$. For this step the loss function is augmented to include a \VICreg{} regularization term; see Subsection~\ref{subsec:vicreg} for more information on the regularization, and Subsection~\ref{subsec:datacompress} for technical details of the data compression. \label{item:finalcompression}}
    
\end{enumerate}
We will use $\textbf{t}$ to denote the output of the final $G^*$ compression step.

\begin{figure*}
    \centering
    \includegraphics[width=\textwidth]{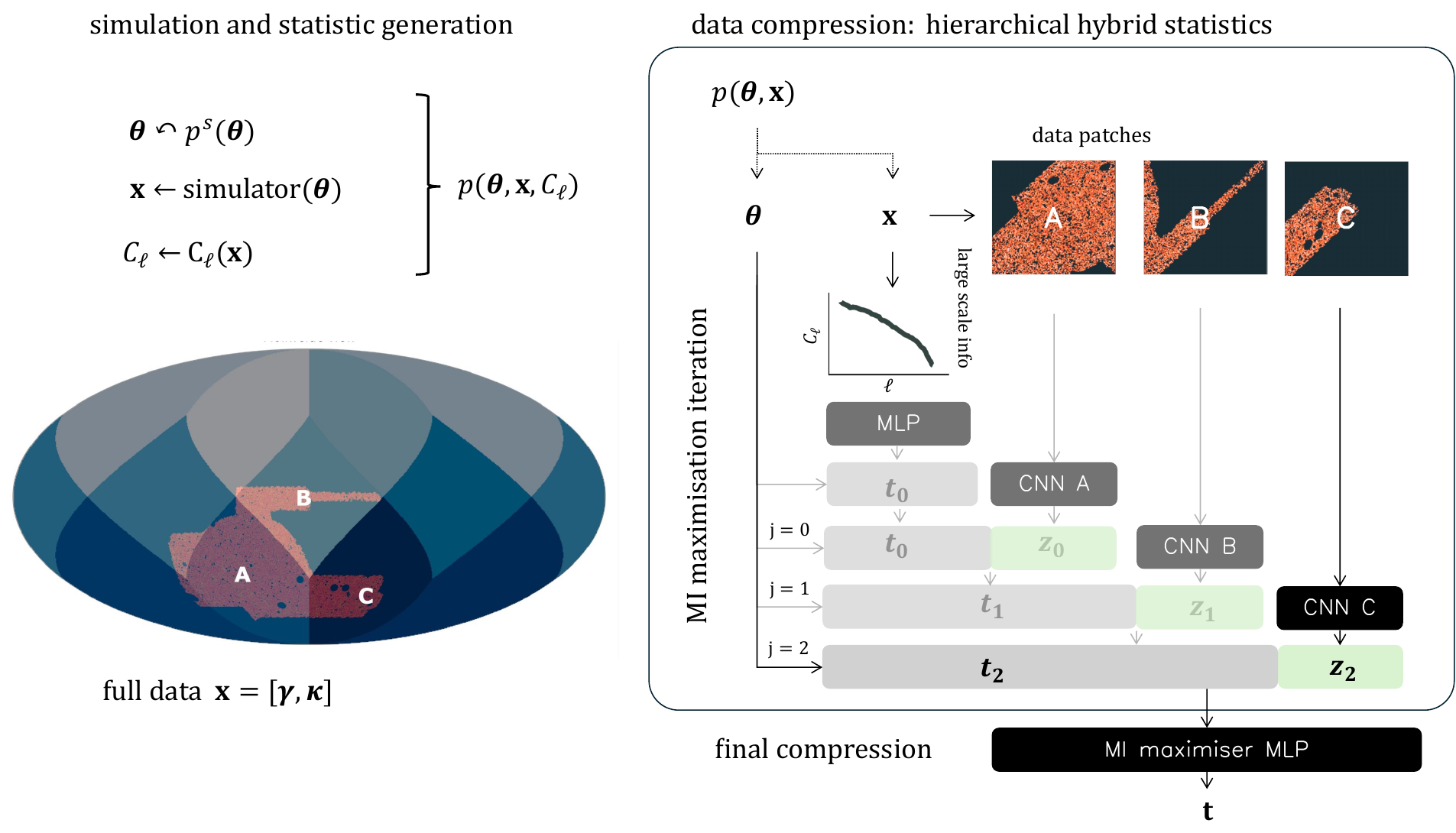}
    \caption{Hierarchical hybrid statistics scheme adapted for DES Y3 data products. The DES footprint is simulated over a simulation sampling distribution in cosmology $\simsampdist(\parvec)$ with nuisance parameters (\textit{left}). Empirical $C_\ell$ vectors are computed from the full shear field ($\gamma$) footprint and compressed via MI maximization to $\dim(t_0)=10$ numbers. The reconstructed convergence field $\kappa$ is split into patches A,B, and C and passed hierarchically through separate CNNs to $\dim(z_j)=4$ numbers to probe small scales and complement existing summaries in sequence, with network weights from previous steps frozen, resulting in 22 summaries per simulation. The scheme is repeated four times with different network initializations, where the concatenated 88 summaries are finally compressed to a summary vector $\textbf{t}$ of seven numbers.} 
    \label{fig:compression-scheme}
\end{figure*}

\subsection{Variance-invariance-covariance regularization} \label{subsec:vicreg}

To regularize the final compression step of our hierarchical scheme (step (v) in Subsection~\ref{subsec:hierarchicalcompression}), we supplement the VMIM loss with variance-invariance-covariance regularization (\VICreg{}; \citealt{VICreg2021}). \VICreg{} operates on batches of embedding vectors and comprises three complementary terms. Given a batch of $n$ compressed summary vectors $T = [\textbf{t}_1, \dots, \textbf{t}_n]$ of dimension $d$, the variance term
\begin{equation}
v(T) =  \avg_{j=1}^{d} \max\!\left(0,\, \gamma - \sqrt{\mathrm{Var}(t^j) + \epsilon}\right)
\end{equation}
is a hinge loss that maintains the standard deviation of each summary dimension above a threshold $\gamma$, preventing the compressor from collapsing dimensions to constant values. The covariance term
\begin{equation}
c(T) = \frac{1}{d} \sum_{i \neq j} [C(T)]^2_{i,j},
\end{equation}
where $C(T)$ is the sample covariance matrix, penalizes off-diagonal correlations and encourages the learned summaries to encode non-redundant information. The invariance term
\begin{equation}
s(T, T') = \frac{1}{n} \sum_{i=1}^{n} \lVert \textbf{t}_i - \textbf{t}'_i \rVert^2
\end{equation}
encourages summaries derived from different realizations of the same parameters $\boldsymbol{\theta}$ to remain close in the summary space. The total regularization loss is
\begin{equation}
\lossfunction{}_{\mathrm{VIC}}(T, T') = \lambda_{\nu}[v(T) + v(T')] + \lambda_{c}[c(T) + c(T')] + \lambda_{s}s(T,T').
\end{equation}
This is added to the VMIM loss $\mathcal{L}$ of equation (\ref{eq:loss_function}) in the final compression only. The variance and covariance regularization weights, $\lambda_{\nu}$ and $\lambda_{c}$ respectively, are set to 0.2, while the invariance regularization weight, $\lambda_{s}$, is set to $0$ (we found that doing otherwise results in summaries that are no better in terms of invariance and that are worse in terms of covariance). The principal benefit of adding the \VICreg{} loss is that the resulting summary space distribution is substantially easier for downstream density estimation to condition on. We discuss this further in Subsection~\ref{subsec:postunblindingchanges}.

\section{Weak gravitational lensing}\label{sec:weaklensing}

In this study we use convergence mass maps inferred from weakly lensed galaxy shapes; below we derive how this quantity arises from the underlying projected matter density field and relate it to weakly lensed galaxy shapes. 

Here we summarize the weak lensing formalism. For further details see \citet{Hu2000} and reviews such as \citet{Bartelmann_2001}, \citet{Munshi}, and \citet{Kilbinger}. The gravitational potential $\Phi$ in a comoving spatial coordinate system $\mathbf{r}$, at time $t$, depends on the matter overdensity $\delta = \delta \rho/\bar{\rho}$ via the Poisson equation
\begin{equation}
\label{eq:poisson}
\nabla^2_r \Phi(t, \mathbf{r}) = \frac{3 \, \Om H_0^2}{2 a(t)} \delta(t, \mathbf{r}) \ ,
\end{equation}
where $a \equiv 1 / (1 +z)$ is the scale factor, $\Om$ is the total matter density today, and $H_0$ is the Hubble constant today. Defining a source plane at comoving distance $\chi_s$ and in direction $\direction{}$ from the observer, we can define the three-dimensional lensing potential (assuming the Born approximation and assuming a flat universe)

\begin{equation}
\label{eq:3Dlensingpotential}
\phi(\direction{}, \chi_s)
= \frac{2}{c^{2}} \int_{0}^{\chi_s} \frac{\chi_s - \chi}{\chi_s\,\chi} \,
\Phi\!\bigl(\direction{}, \chi\bigr) \, \mathd \chi
\end{equation}

\noindent in terms of comoving distance $\chi$. We then integrate over the line of sight, weighting the lensing potential by a normalized redshift distribution of source planes $n(z(\chi_s))$, to get the angular lensing potential
\begin{equation}
\label{eq:2Dlensingpotential}
\phi(\direction{}) = \int n(z(\chi_s)) \phi(\direction{}, \chi_s) \, \mathd \chi_s \, .
\end{equation}

The potential is defined on the celestial sphere, so it is useful to represent it and its second derivatives via the formalism of spin-weighted functions and the associated differential operators $\eth$ and $\bar{\eth}$; see \cite{castro2005weak} for details. The potential is scalar, i.e. of spin weight 0. From its Hessian matrix we derive:
\begin{itemize}[wide]
\item{
the \textit{convergence} $\kappa$, of spin weight 0, (one half of) the trace of the Hessian (and hence half the Laplacian of $\phi$), representing an isotropic change in area due to lensing:
\begin{equation}
\label{eq:kappaphi}
\kappa = \frac{1}{4} (\eth \bar{\eth} + \bar{\eth} \eth) \phi \, ;
\end{equation}
}
\item{
the \textit{shear} $\gamma = \gamma_1 + \imagunit \gamma_2$, of spin weight 2, the traceless part of the Hessian, representing an anisotropic area-preserving shape change due to lensing:
\begin{equation}
\label{eq:gammaphi}
\gamma = \frac{1}{2} \eth \eth \phi \, .
\end{equation}
}
\end{itemize}

Equations (\ref{eq:poisson}) and (\ref{eq:kappaphi}), both involving the Laplace operator, allow us to rewrite the lens equation (\ref{eq:2Dlensingpotential}) as 

\begin{multline}
\kappa(\direction{}) = \\ 
\frac{3\,\Om H_0^2}{2c^2}
\int_{0}^{\infty} n(z(\chi_s))
\int_{0}^{\chi_s} \frac{\chi \, (\chi_s - \chi)}{\chi_s} \,
\frac{\delta\bigl(\direction{}, \chi \bigr)}{a(\chi)} \, \mathd \chi \, \mathd \chi_s
\label{eq:convergence}
\end{multline}
i.e. convergence is a projected weighted-average matter overdensity.

If we represent $\phi$, $\kappa$, and $\gamma$ in terms of the basis given by spherical harmonic functions of the appropriate spin weight then equations (\ref{eq:kappaphi}) and (\ref{eq:gammaphi}) become

\begin{equation}
\label{eq:kappagamma_alm_new}
\hat{\kappa}_{\ell m} = -\frac{1}{2}\,\ell(\ell+1)\,\hat{\phi}_{\ell m}, 
\quad
\hat{\gamma}_{\ell m} = \frac{1}{2}\sqrt{(\ell-1)\,\ell\,(\ell+1)\,(\ell+2)}\,\hat{\phi}_{\ell m},
\end{equation}
respectively, whence

\begin{equation}
\label{eq:kappagamma_alm_relation}
\hat{\kappa}_{\ell m}
= - \sqrt{\frac{\ell(\ell+1)}{(\ell-1)(\ell+2)}}\,\hat{\gamma}_{\ell m}.
\end{equation}
Now $\hat{\gamma}_{\ell m}$ is defined only for $\ell \ge 2$, and hence equation~(\ref{eq:kappagamma_alm_relation}) does not constrain $\hat{\kappa}_{\ell m}$ for $\ell = 0, 1$; this is the \textit{mass sheet degeneracy}, which we resolve by setting $\hat{\kappa}_{\ell m} = 0$ for $\ell = 0, 1$. 

The shear is derived from the shapes of galaxies and includes the distortions from the intervening mass along the projected line of sight as well as other contributions not outlined here (such as shape noise, intrinsic alignments, and limited survey sky area). Since for any single galaxy the unlensed shape is unknown and weak lensing has only a tiny effect on an individual galaxy's shape, photometric galaxy surveys average over large numbers of source galaxies aggregated into pixels (on the celestial sphere) and tomographic bins (in redshift), such that individual variations in intrinsic galaxy shape average out and a weak lensing shear signal emerges. Equation (\ref{eq:kappagamma_alm_relation}) allows the reconstruction of the convergence mass map from shear data; this is \textit{Kaiser-Squires reconstruction} \citep{KS_1993}.

\section{Simulations and Dark Energy Survey data}\label{sec:simsdata}
\subsection{Gower Street simulations}
In this work we use the Gower Street (v1) suite of simulations, introduced in \citet{jeffrey2024dark}.

The simulation suite is designed for DES and \textit{Euclid} survey SBI analyses. The suite consists of 791 simulations, sampling a 7-dimensional $\nu w$CDM cosmology defined by: the matter density $\Om$, the dark energy equation of state $w$, the amplitude of matter fluctuations $\sigma_8$, the physical baryon density $\Omega_{\rm b}h^2$, the sum of neutrino masses $m_\nu$, the dimensionless Hubble parameter $h$, and the scalar spectral index $n_s$. The simulations are run using the \pkdgrav{} code \citep{potter2017pkdgrav3}. The distribution of parameters is different from the priors imposed in the inference analysis, which motivates our choice of NLE parameterization, as opposed to NPE, in the density estimation and posterior coverage steps.

The Gower Street suite uses $N$-body gravity-only simulations in a full sky box of size $L = 1250\,h^{-1}\,\mathrm{Mpc}$ with number of particles per side $N = 1080$ from an initial redshift $z_0 = 49$ to a final zero redshift. The initial conditions are generated using second-order Lagrangian Perturbation Theory (2LPT), which are initialized randomly in each simulation to preserve cosmic variance in the suite. The code calculates the intersections (in spacetime) of the observer's lightcone with the worldlines of each of the $N^3$ particles; these intersection events are then binned both temporally (into redshift bins) and spatially (into pixels on the sky). To trace the observer's light cone back to the starting redshift, the simulation box is repeated $(2 \times 10)^3$ times to form a `super-box' with the observer at the centre (although note that $(2 \times 3)^3$ replicated boxes will reach back to $z \sim 1.5$, sufficient for DES Year 3 modelling).

The sampled $\nu w$CDM parameters are summarized in Table~\ref{tab:gowerstreetprior}. In this table, $\mathcal{N}(\mu, \sigma)$ denotes a normal distribution of mean $\mu$ and standard deviation $\sigma$, $\mathcal{U}[a, b]$ denotes a uniform distribution between $a$ and $b$, and $\texttt{\textrm{ActiveLearning}}[a, b]$ denotes a distribution supported on $[a,b]$ from which we use active learning to sample. See Fig.~4 of \cite{jeffrey2024dark} for a corner plot of the full joint distribution. We comment:

\begin{itemize}[wide]
    \item $\Om$ and $\sigma_8$ are the parameters most tightly constrained by weak lensing, with the two often combined into $S_8\equiv \sigma_8(\Om/0.3)^{0.5}$; therefore, an active learning approach, based on DES Year 3 posteriors, is used for these input parameters to ensure a high concentration of simulations in the anticipated inference region.
    \item $n_s$ is chosen from Planck \citep{aghanim2020planck}, but with the standard deviation boosted by 1.5. 
    \item $h$ is distributed with the mean half-way between Planck \citep{aghanim2020planck} and SH0ES \citep{Riess_2022}, and with a one standard deviation interval wide enough to encompass the two standard deviation intervals of both Planck and SH0ES. 
    \item $\Omega_{\rm b}h^2$ is consistent with Planck \citep{aghanim2020planck}. 
    \item $w$ is driven by the choice of prior in the original DES Year 3 SBI weak lensing analysis that this work is based on. However, in the validation runs, 64 simulations have values outside this prior range; these runs are not discarded as they smooth the hard boundaries at the edges of the $w$ prior. 
    \item $m_\nu$ mirrors similar choices to the DES Y3 analysis. For the first 192 simulations the treatment of neutrinos is more simplistic and only modelled in the initial conditions at a fixed mass of 0.06. The remaining runs beyond verification use the \concept{} code to model the effect of neutrinos \citep{tram2019}.
\end{itemize}

\begin{table}
\caption {The parameter distributions of the Gower Street simulations.}
\centering
\begin{tabular}{ c c  }
\toprule
\textbf{Parameter} & \textbf{Simulation Sampling Distribution} \\
\midrule
$\Om$   & $\texttt{\textrm{ActiveLearning}}[0.15, 0.49]$ \\
$\sigma_8$   & $\texttt{\textrm{ActiveLearning}}[0.5, 1.0]$ \\
$h$          & $\mathcal{N}(0.7022, 0.0245)$ \\
$n_s$        & $\mathcal{N}(0.9649, 0.0063)$ \\
$\Omega_{\rm b}h^2$ & $\mathcal{N}(0.02237, 0.00015)$ \\
$w$          & $\mathcal{U}[-1, -\tfrac{1}{3}]$ \\
$\ln(m_{\nu})$    & $\mathcal{U}[\ln(0.06), \ln(0.14)]$ \\
\bottomrule
\end{tabular}
\label{tab:gowerstreetprior}
\end{table}

\begin{figure}
    \centering
    \includegraphics[width=0.8\linewidth]{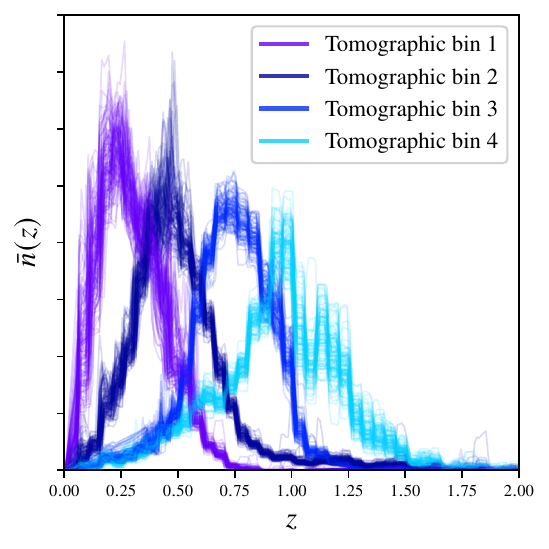}
    \caption{Samples from $p(\bar{n}(z) \cond x_{\rm phot})$ of the DES Y3 footprint redshift distribution from \texttt{SOMPZ} \protect\citep{y3-sompz}.}
    \label{fig:nz_sompz_samples}
\end{figure}

\subsection{DES Y3 data}

The Dark Energy Survey (DES) is a multi-band imaging program carried out with the Dark Energy Camera on the Blanco 4-m telescope, designed to map roughly $5000~\mathrm{deg}^2$ of the southern sky in the $grizY$ filters for studies of cosmic acceleration and structure formation.  The data from the first three years of DES operations are assembled in the Y3\,\textsc{gold} photometric data set \citep{Sevilla-Noarbe_2021}, which provides uniformly calibrated photometry and masks, covering an area of $\sim 5000~\mathrm{deg}^2$.  Cosmological weak lensing and galaxy–galaxy lensing analyses draw on a subset of this parent catalogue for which accurate galaxy shapes, photometric redshifts, and associated selection information are available.

The main input dataset to this analysis is the DES Year 3 weak lensing shape catalogue \citep{y3-shapecatalog}, constructed from Y3\,\textsc{gold} and containing shape measurements for $100{,}204{,}026$ galaxies over $4143~\mathrm{deg}^2$ of the footprint, using multi-epoch imaging in the $riz$ bands.  Galaxy ellipticities are measured with the \textsc{metacalibration} algorithm \citep{HuffMcal2017}, which self-calibrates shear response by applying small artificial shears to the images and remeasuring shapes, delivering a catalogue with effective source number density $n_{\rm eff} = 5.59~\mathrm{arcmin}^{-2}$ and per-component shape noise $\sigma_e = 0.261$. Extensive null tests are applied to verify that residual systematics from point-spread-function modelling, blending, selection effects, and other observational conditions are below the statistical error budget, making the DES Y3 shape catalogue a key input for the suite of Y3 cosmology analyses based on two-point and higher-order weak lensing statistics.

\begin{figure}
    \centering
    \includegraphics[width=0.8\linewidth]{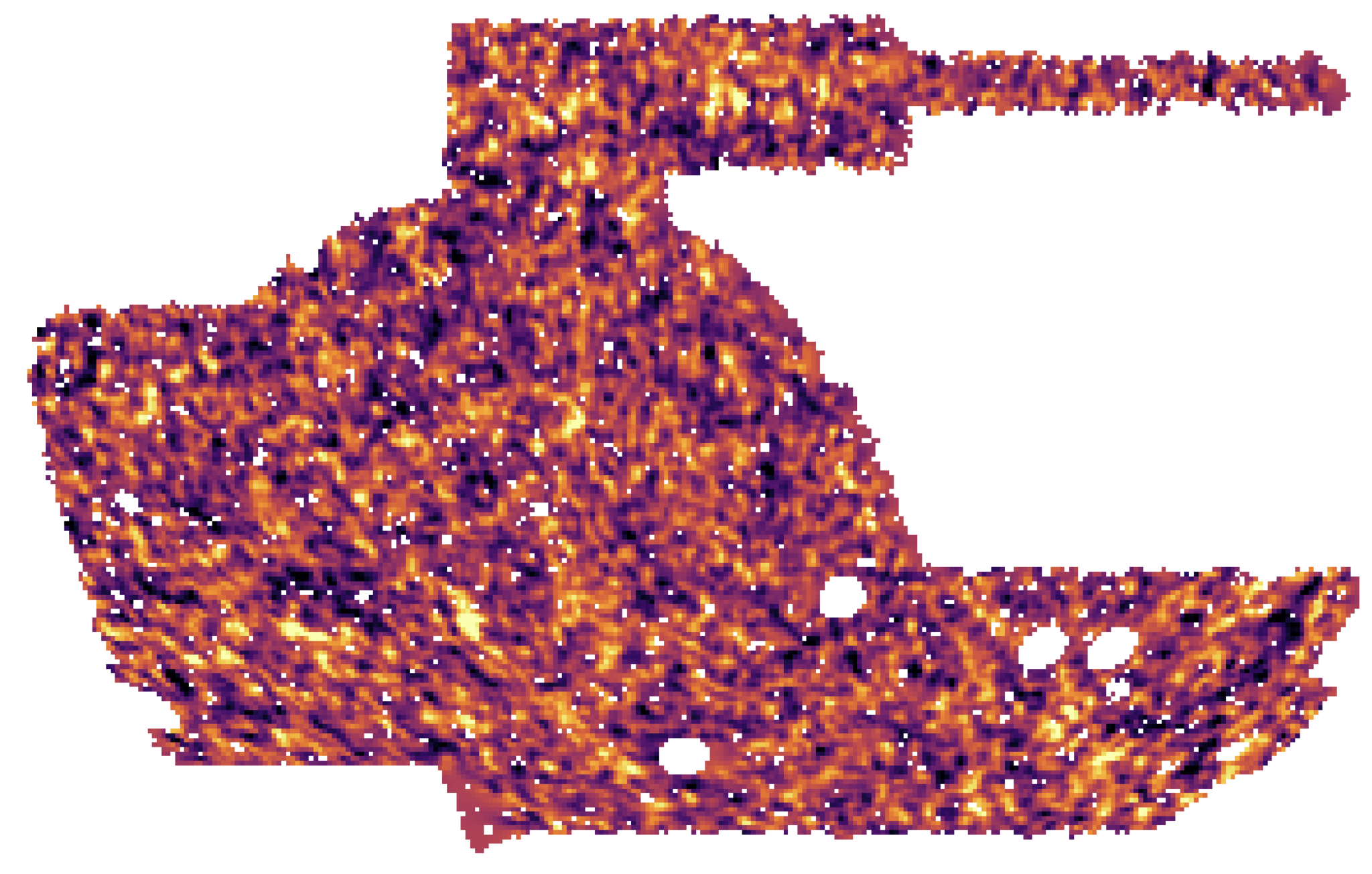}
    \caption{The DES Y3 full projected mass map at a \healpix{} nside resolution of $1024$, which differs from the analysis resolution of $512$. This convergence map is produced by Kaiser-Squires reconstruction, as detailed by \protect\cite*{y3-massmapping}, using all the source tomographic bins from the \textsc{metacalibration} catalogue.}
    \label{fig:y3_footprint}
\end{figure}

The shape catalogue is processed in \cite*{y3-massmapping} using the Kaiser-Squires reconstruction to give the input convergence maps (Fig.~\ref{fig:y3_footprint}) that we use for the main inference.

\subsection{Forward modelling}\label{subsec:forwardmodelling}

We follow the same methodology as \cite{jeffrey2024dark} to forward model from Gower Street simulations to realistic DES Year 3 mock shear maps. The modelling choices are summarized here; for more detail see \cite{gatti_dark_2024}. From the Gower Street simulations, we have $\sim 100$ shells along the observer's light cone, spaced equally in comoving distance such that for shell $s$, we can create a pixelized matter overdensity $\delta_\textnormal{shell}(\boldsymbol{\theta}, \chi_s)$ as a \healpix{} map. 

From these overdensity shells we can construct a noiseless convergence map using equation (\ref{eq:convergence}); this process is implemented in the BornRaytrace code\footnote{\url{https://github.com/NiallJeffrey/BornRaytrace}}. For each $i^{\textrm{th}}$ source tomographic bin we use equation (\ref{eq:convergence}) to integrate over the overdensity shells to get a noiseless convergence field $\kappa^i(\boldsymbol{\theta})$. We then convert to a noiseless projected shear map $\gamma^i(\boldsymbol{\theta})$; for this we use the inverse Kaiser-Squires transform given by equation (\ref{eq:kappagamma_alm_relation}).

For each full sky Gower Street simulation, we partition the sky into four pseudo-independent DES footprints. For each DES footprint we generate independent samples of survey and systematics noise. We model photometric redshift uncertainty using the \hyperrank \ method from \citet{y3-hyperrank}, which generates redshift distributions that are consistent with the data (examples shown in Fig.~\ref{fig:nz_sompz_samples}). We draw from this distribution of redshifts for each DES footprint cutout. 

Galaxies nearby to each other can have correlated shapes as the result of local tidal fields and galaxy formation processes, independent of weak lensing effects; these local correlations are called intrinsic alignments. We model intrinsic alignments using the NLA model \citep{hirata_seljak}. We can model the intrinsic alignment contribution to the convergence map:

\begin{equation}
\label{eq:NLA}
\kappa_{\textrm{IA}} (\direction{},z) = - A_{\textrm{IA}} C_1 \rho_{\textrm{crit}}  \frac{\Om}{D(z)}  \Big( \frac{1+z}{1+z_0} \Big)^{\eta_{\textrm{IA}}}  \ \delta(\direction{},z) \, .
\end{equation}
We fix $z_0=0.62$ and $C_1 = 5\times 10^{-14}\,M_\odot\,h^{-2}\,\mathrm{Mpc}^2$ \citep[as found by][]{Bridle2007}. The intrinsic alignment amplitude $A_{\textrm{IA}}$ and the redshift evolution parameter $\eta_{\textrm{IA}}$ are treated as nuisance parameters with conservative priors in this analysis, as seen in Table~\ref{tab:priors}. $D(z)$ is the linear growth factor.

The shear measurement $\gamma$ is typically biased by a combination of noise, shape fitting, PSF, and blending effects. This bias can be modelled by a multiplicative shear bias factor so that $\gamma=(1+m)\gamma_{\textnormal{true}}$; such a factor has been extensively characterized by \citet{MacCrann_2021}. We specify $m$ (one per tomographic bin) using the prior shown in Table~\ref{tab:priors}.

We also model the effects of source clustering as \citet{source_clustering} observed that higher-order statistics are sensitive to the source galaxies tracing the underlying matter distribution. Instead of characterizing the sources by an isotropic redshift distribution, we model the angular distribution with a linear galaxy bias model

\begin{equation}
    \label{eq:sourceclustering}
    n(z,\direction{}) \propto n(z)(1+b_g\delta(z,\direction{})).
\end{equation}

In addition to the shear signal, we also model the shape noise contribution to the observed shear from the DES Year 3 catalogue. We randomly rotate DES galaxies with ellipticity $e_g$ and catalogue weight $w_g$ to erase the cosmology signal. We model the shear signal in pixel $p$ of our \healpix{} maps by summing the contributions of our simulation shells $s$ per tomographic bin

\begin{multline}
\label{eq:sc_pixel}
\gamma(p) = \frac{\sum_s \bar{n}(s) [1 + b_g \delta(p, s)] (1 + m) [\gamma(p, s)+\gamma_{\textrm{IA}}(p, s)]}{\sum_s \bar{n}(s) [1 + b_g \delta(p, s)]}  + \\
\left(\frac{\sum_s \bar{n}(s)}{\sum_s \bar{n}(s) \left[1 + b_g \delta(p, s)\right]}\right)^{1/2} F(p) \, \frac{\sum_g w_g e_g}{\sum_g w_g},
\end{multline}
where the first term is the signal (including intrinsic alignments) and the second term is the shape noise. We model the shape noise using the DES Y3 catalogue, which already contains signal from the source clustering seen in Y3. Therefore following \citet{source_clustering} we use a scale factor $F(p)$ which scales the even moments of the map shape noise to mitigate modelling the source clustering already contained in the Y3 data:

\begin{equation}
    F(p) = A\sqrt{1-B \sigma_{e}^2(p)}
\end{equation}
where $\sigma_{e}^2(p)$ is the variance of the shape noise in pixel $p$
and where $A = [0.97, 0.985, 0.990, 0.995]$ and $B = [0.1, 0.05, 0.035, 0.035]$ are constants (one $A$ and one $B$ per tomographic bin).

We run this process for all DES Y3 footprints (each simulation provides four such cutouts), for all four tomographic bins, producing $3{,}164$ independent DES Y3 shear maps for a single pass of the forward model on the Gower Street suite. The survey mask is modelled consistently with the Y3 data. We run the forward model four times sampling randomly from the forward model priors in each run and creating noise realizations; this gives a total of $12{,}656$ mocks for network compression training and NDE training. We hold out one noise realization from the map-level compression, to be used for training and validating our NDEs for the final SBI analysis.

\subsection{Power spectra, patches, and scale cuts}

\subsubsection{Maps and patches} The maps are prepared by modelling the DES Y3 survey mask in \healpix{} \citep{gorski2005} format at a resolution of \nside{}$512$ for the full sky shear fields detailed in Subsection~\ref{subsec:forwardmodelling}. The shear fields are then converted to convergence E- and B-mode fields using the Kaiser-Squires reconstruction outlined in Section~\ref{sec:weaklensing}. We note that the Kaiser-Squires reconstruction can induce artifacts such as mode-mixing and mask edge effects into the derived convergence map; however, as we are forward modelling the actual data with the same KS transformation as the simulations, this does not result in a source of bias in this analysis. At a resolution of \nside{}$512$, the pixel window function results in a smooth suppression of power at $\ell > 1024$ with non-zero support diminishing at $\ell \sim 1500$. During the construction of these maps we enforce a hard cut of $\ell<1024$ in harmonic space to minimize the risk of bias due to baryonic effects present at small scales. 

Although the \healpix{} pixelation is defined on the sphere and there exist neural network architectures that can operate on spherical geometries (see \citealt{deepsphere_iclr} and \citealt{DISCO} for existing implementations), we perform this analysis using the same \textit{patching} scheme as the existing map-level compression in \cite{jeffrey2024dark}. Here the DES Y3 footprint is split into three patches: A, B, and C (see Fig.~\ref{fig:compression-scheme}). Each patch is the set of pixels encompassed by the superpixel of an \nside{1} resolution map. The patches A, B, and C capture most of the DES Y3 footprint aside from a small amount that is discarded; the resulting patches are of resolution $512 \times 512$ pixels. Due to the patches being treated as flat in this CNN implementation, this patch transformation will induce some distortions away from spherical geometry. While it may be a suboptimal representation of the map, we are applying this transformation consistently with the actual data and so the transformation will not bias the analysis. The patch size is a balance between distorting the spherical geometry and including larger scales. The hybrid statistics pipeline deals with this compromise well as the larger scales are inherently learned conditionally from the two-point data, whereas excess information at smaller scales is available at the patch level.

\subsubsection{Power spectra} From the shear maps described at the end of Subsection~\ref{subsec:forwardmodelling}, we implement the following procedure for estimating the weak lensing power spectrum $C(\ell)$, defined in spherical harmonic space as 
\begin{equation}
  \langle a_{\ell m}^{} a_{\ell' m'}^* \rangle =  C_\ell\delta_{m m'} \delta_{\ell \ell'} \ ,
\end{equation}
and assuming statistical isotropy 
\begin{equation}
    \hat{C}(\ell) = \frac{1}{2 \ell +1} \sum_{m=-\ell}^{\ell} \left|a_{\ell m}\right|^2 \ \ .
\end{equation}
 The power spectrum is calculated on the shear field, using the same pseudo-$C_\ell$ code used in the previous DES Y3 analysis \citep[see][]{Doux_2022}. We decompose the shear angular power spectra into the autocorrelation of the E- and B-modes ($C_\ell^{EE}$ and $C_\ell^{BB}$) as our data vector for the two-point input to the hybrid statistics pipeline. The binning and band powers in $\ell$ match those in \citet{Doux_2022} for the auto- and cross correlation of all the source tomographic bins. We exclude from the data vector any bin that contains band power of $\ell > 1024$, to remain consistent with the scale cuts at the map level. In Subsection~\ref{subsec:systematics} we validate that our results are insensitive to unmodelled baryonic feedback effects at the scales we include.

\section{DES Y3 analysis pipeline}\label{sec:compression}

\subsection{Analysis priors}
The priors used for cosmological inference are shown in the top section of Table~\ref{tab:priors}. We enforce wide uniform priors on the lensing parameters $\{ \Om, S_8 \}$, which are different from the more concentrated Gower Street active learning sampling distribution in this plane but share the same support. The priors differ slightly from the fiducial DES $3\times2$pt analysis to reflect the range spanned by the Gower Street simulations. The lower bound $w\geq-1$ excludes phantom dark energy from our analysis. The remaining parameters 
are not inferred, but are implicitly marginalized in the analysis such that their uncertainty is represented in our target parameters.

\begin{table}
\caption {Priors used in this analysis \citep[following][]{jeffrey2024dark}.}
\centering
\begin{tabular}{ c c }
\toprule
\textbf{Parameter} & \textbf{Prior} \\
\midrule
$\Om$ & $\mathcal{U}[0.15, 0.52]$ \\
$S_8$ & $\mathcal{U}[0.5, 1.0]$ \\ 
$w$ & $\mathcal{U}[-1, -\tfrac{1}{3}]$  \\
\midrule
$n_s$ & $\mathcal{N}(0.9649, 0.0063)$   \\
$h$ & $\mathcal{N}(0.7022, 0.0245)$ \\
$\Omega_{\rm b}h^2$ &  $ \mathcal{N}(0.02237, 0.00015)$\\
$\ln(m_{\nu}) $& $ \mathcal{U}[\ln(0.06), \ln(0.14)]$ \\
\midrule
$A_{\textrm{IA}}$ & $\mathcal{U}[-3, 3]$  \\
$\eta_{\textrm{IA}}$ & $\mathcal{U}[-5, 5]$ \\
$m_{1}$ & $\mathcal{N}(-0.0063,0.0091)$ \\
$m_{2}$ & $\mathcal{N}( -0.0198,0.0078)$ \\
$m_{3}$ & $\mathcal{N}( -0.0241,0.0076)$ \\
$m_{4}$ & $\mathcal{N}(-0.0369, 0.0076)$ \\
$\bar{n}_i(z)$ & $p_{\hyperrank{}}(\bar{n}_i(z) \cond x_{\rm phot})$  \\ 
\bottomrule
\end{tabular}
\label{tab:priors}
\end{table}

\subsection{DES Y3 data compression}\label{subsec:datacompress}

Our objective is to obtain joint constraints on the three-dimensional weak lensing--dark energy parameter set $(\Om, S_8, w)$.

For our explicit implementation of the procedure, we choose to parameterize the mutual information maximizer surrogate $q(\parvec \cond \cdot)$ with a simple mixture density network \citep[MDN;][]{bishop_mdn_1994} comprised of a mixture of four Gaussians and a single ReLU-activated hidden layer of width 100. We found empirically that simpler MDN configurations performed better, that is, the gradient descent optimization is balanced between changes in the compression networks $F_j$ and NDE $q_j$ at each stage of the procedure, as opposed to dominated by the density estimation operation. We detail each stage of the compression. See Fig.~\ref{fig:compression-scheme} for an illustration.

\subsubsect{Initial $C_\ell$ compression} The E- and B-mode $C_\ell$ data vector with appropriate cuts has size $440$. We use the fully connected network employed by \cite{jeffrey2024dark}, with ten hidden layers, each with 256 nodes and an embedded layer normalization with a ReLU activation function at each layer output, producing an output summary size of $\dim(\textbf{t}_0)=10$. We choose this summary dimensionality to allow for a large enough bottleneck for information to propagate from summaries to the NDE $q(\parvec \cond \cdot )$. Each input vector's position was preprocessed by transforming its distribution in the training data to zero mean and unit standard deviation. During training, the input dataset of power spectra and map pixels is augmented with a small amount of additive Gaussian noise at each epoch ($\mu = 0, \sigma = \sigma_{x_i} \times 10^{-3}$) to improve generalization. 

\subsubsect{Patch compression} Here we use the CNN architecture used in \cite{jeffrey2024dark} to compress each patch down to $\dim(\textbf{z}_{\rm patch})=4$ numbers to accompany the $C_\ell$ summaries, yielding a final concatenated summary vector of $10 + 3 \times 4 = 22$ numbers. The inputs to each patch network are the E- and B-modes of each map across the four source tomographic bins, resulting in eight CNN channels of $512 \times 512$ pixels. Each pixel from every channel is transformed to zero mean and unit standard deviation (matching the treatment of the $C_\ell$ data vector in the previous section). At each epoch, every pixel in the data vector is augmented with Gaussian noise.  On each hybrid statistic iteration $j$, we initialize each network anew for each data product's compression. We operate on patches A, B, and C in sequence; other permutations might yield different results, as the hybrid formalism is not permutation-invariant. We repeat the hybrid compression four separate times with different Lecun normal initializations, resulting in an ensemble of 88 hybrid summary vectors. 

\subsubsect{Final compression} We compress the final accumulated hybrid statistic vector of length 88 down to $\dim(\textbf{t})=7$ numbers using the `VMIM plus \VICreg{}' loss as described in Subsection~\ref{subsec:vicreg}. These statistics are then used for NLE density estimation. Overfitting is a common problem in CNNs; it is a particular problem in SBI as overfitting results in a network that does not learn informative field-level features and instead memorizes patterns in the training dataset, violating the data processing inequality. We perform SBI on a noise realization held out from the training. Therefore, overfitting will not bias our inference if it does occur; it will instead simply make our posteriors suboptimal (i.e. wider). To mitigate the effects of overfitting, we stop early if the validation loss does not improve within 100 epochs. 

The hybrid statistics network was trained on $9{,}264$ DES Y3 mock maps (comprising three independent noise realizations of our forward modelling pipeline with $3{,}088$ mocks per noise realization). The validation loss from which the early stopping criterion is calculated is run on a further independent noise realization of the forward model. We hold an independent noise realization of $3{,}088$ mocks completely blind to the compression networks, specifically for training and validating the NDEs for estimating the likelihood we use in our analysis.

\subsubsection{Comparison to previous compression scheme} For data compression, \cite{jeffrey2024dark} optimized a mean-square error (MSE) objective with respect to each parameter $\theta_j$
and data product $t_i$ \textit{separately}:

\begin{equation}
    \hat{\theta}^*_{ij} = \min_{f_i} \, \avg_{b=1}^{N} \, (f_i(t_i^b) - \theta_j^b)^2 \, .
\end{equation}
These are then aggregated into parameter estimates using a weighted average over $n_d$ data products
\begin{equation}
    \hat{\theta}_j = \avg_{i=1}^{n_d} \, \hat{\theta}^*_{ij} w_i,
\end{equation}
where the weights are computed from the best validation loss per data product. The MSE criterion minimized over the parameter prior converges to the posterior mean and can be locally made equivalent to MOPED compression \citep{moped, momentnets, alsing2018_general}. This compression scheme is optimal in the case that the parameters being inferred have a Gaussian distribution.

\subsection{NLE implementation}\label{subsec:nde_details}
To parameterize the likelihood of our summary vector $p(\textbf{t} \cond \parvec)$ we employ an ensemble of eight MAFs, with hidden size 200 and 6, 7, 8, or 9 transformations, within the \texttt{ltu-ili} package \citep{ho2024ltuili}.
Each network of the ensemble is trained independently from different network parameter initial values. The NDEs are trained with 30 per cent of the simulations held out for validation. In contrast to the neural compression step, overfitting during NDE training can result in bias in the inferred likelihood. The training has an early stopping threshold implemented on the held-out validation dataset to prevent overfitting; as well, a coverage test (see Subsection~\ref{subsec:coverage}) verifies the results of the NDEs.

To avoid potential artifacts from training on samples with a hard boundary \citep{cornish2020relaxing,tirapongprasert2026learningedgetaileduniformsampling}, we use a training set that extends into the phantom regime, $-1.05<w<-0.3$, and apply the hard prior separately later.
We ensure that posteriors and coverage tests are insensitive to the chosen $w$ cutoff.

To sample the posterior distribution we use an affine sampler MCMC \citep[see e.g.][]{Foreman_Mackey_2013} to evaluate the learned, amortized summary log-likelihood over the specified cosmological prior.

\section{Inference validation}\label{sec:validation}
This section describes three inference validation tests: an NDE coverage test, a systematics test, and a posterior predictive test. We also describe our blinding strategy.

Small amendments were made to the algorithm following unblinding; see Subsection~\ref{subsec:postunblindingchanges} for details. The inference validation tests were performed both on the original pre-unblinding code as well as on the amended post-unblinding code; the results described below are for the post-unblinding algorithm. Except for the coverage test, the pre-unblinding results were only marginally different.

\subsection{Neural likelihood and coverage tests}
\label{subsec:coverage}

We perform a coverage test to validate the density estimation scheme; such a test checks whether credible intervals contain the expected fraction of simulations.
Specifically, we:
\begin{enumerate}[label=\alph*), wide]
\item{sample $\parvec_{\rm test}$ from an appropriate prior, \label{item:samplefromprior}}
\item{run our forward model and compressor to get corresponding compressed data $\textbf{t}_{\rm test}$,} \item{ perform inference to obtain a posterior $p(\parvec \cond \textbf{t}_{\rm test})$ -- in this step we use the same prior as in \ref{item:samplefromprior}, \label{item:performinference}}
\item{calculate a credible region (say a 68 per cent credible region) from this posterior,}
\item{ determine if $\parvec_{\rm test}$ lies in this credible region (expecting an affirmative answer 68 per cent of the time).}
\end{enumerate}
This procedure can be repeated with many $\parvec_{\rm test}$ drawn from the prior, from which the \textit{expected} coverage can be computed, enhancing confidence in the estimation of the true posterior. We use the generalized coverage (\TARP{}) algorithm of \citet{lemos2023samplingbasedaccuracytestingposterior} to check coverage in the joint $\{\Om, S_8, w \}$ dimensions.

Which prior (call it $\tarpprior{}$) should we use in steps \ref{item:samplefromprior} and \ref{item:performinference} of the coverage test? The obvious choice is the simulation sampling distribution $\simsampdist$, as our simulation suite naturally gives us samples from $\simsampdist$ together with forward model runs for these samples (and note that getting additional forward model runs would be prohibitively expensive). However, we also require $\tarpprior{}$ to be available analytically---this is because the \TARP{} algorithm requires us to provide the posterior $p(\parvec \cond \textbf{t}_{\rm test})$ (from step \ref{item:performinference}) as a set of samples, and the MCMC process for generating such samples needs to be able to calculate the prior density for arbitrary $\parvec$. Alas $\simsampdist$ is not available analytically (due to the active learning distributions). The solution is to define $\tarpprior$ to be a simple analytic model (a truncated Gaussian) that is a close fit to $\simsampdist$, and to draw samples of this by rejection sampling on the set of simulation parameters.

The coverage test requires that for each sample $\parvec_{\rm test}$, we rerun our analysis to derive the posterior $p(\parvec \cond \textbf{t}_{\rm test})$. When doing this, we naturally exclude the simulation corresponding to $\parvec_{\rm test}$ from the NDE likelihood estimation training step to provide an independent coverage test for each simulation. This test is valid as long as we verify that the likelihood learned by an NDE trained without a given cosmology, evaluated on that cosmology's summaries, is indistinguishable from the likelihood learned when the cosmology is included in training. We expect this to hold, as our trained NDEs show good convergence and no signs of overfitting. The suite used to test coverage is a held-out run of the forward model described in Section~\ref{sec:simsdata}, for all of the 791 cosmologies and all four of their pseudo-independent sky cutouts; each simulation therefore provides four posteriors for \TARP{}.

\begin{figure}
    \centering
    \includegraphics[width=0.75\linewidth]{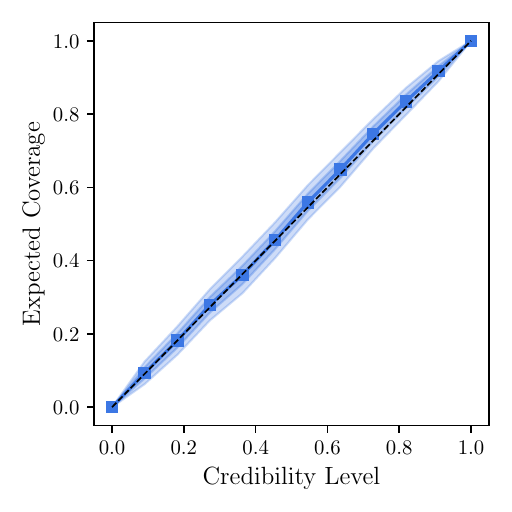}
    \caption{Coverage test result (using \TARP{}; \citealt{lemos2023samplingbasedaccuracytestingposterior})
    to validate the density estimation scheme for hybrid statistics. Using repeated mock data parameter inference, the fraction of true values in the appropriate credible intervals matches the expected fraction. The figure shows the result for all three patches and $C_\ell$ compression. The shaded regions show 1- and 2-$\sigma$ contours (the standard deviation $\sigma$ being calculated via bootstrap resampling).}
    \label{fig:coverage}
\end{figure}

At this point we have the required inputs to \TARP{}:
\begin{itemize}[wide]
    \item{for step a), a set of parameter samples (i.e. samples of $\tarpprior{}$);}
    \item{for step c), a (corresponding) set of posteriors (each posterior itself expressed as a set of samples).}
\end{itemize}
\TARP{} then estimates the coverage probabilities in the $\{\Om, S_8, w \}$ parameter space. We display the coverage test results for the full hybrid statistic vector in Fig.~\ref{fig:coverage}; bootstrap resampling of the \TARP{} calculation allows us to construct error bars on the coverage probabilities (shaded areas in the figure). The excellent results give confidence that the posterior distribution does indeed represent the true underlying parameter distribution given the observed data.

\subsection{Systematics and robustness testing} \label{subsec:systematics}
We test that the statistics used in the pipeline are robust to sources of systematic error or model misspecification in the training set distribution.

\begin{figure}
    \centering
    \includegraphics[width=\linewidth]{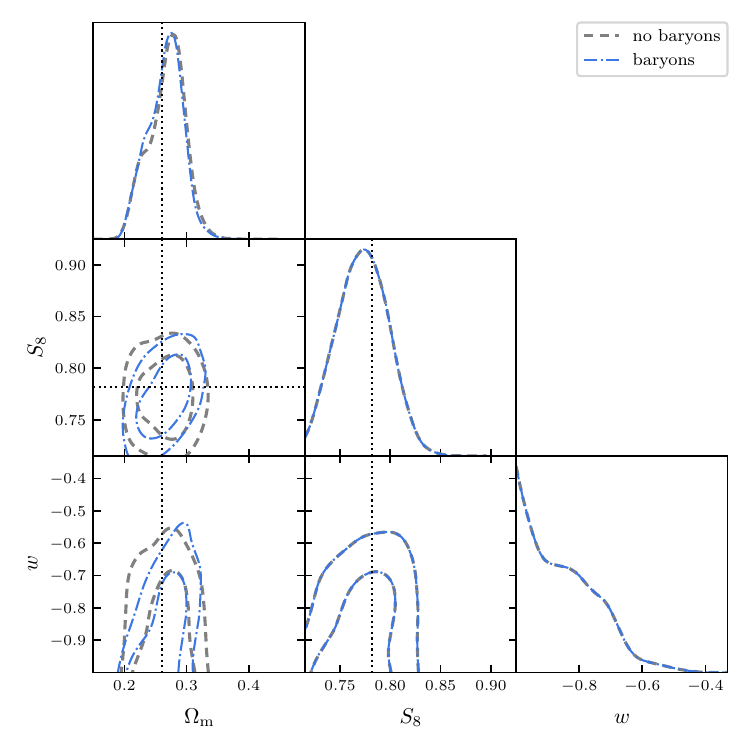}
    \caption{Baryonic feedback systematic error test with the \texttt{CosmoGridV1} simulations. The mean of the inferred marginal posterior distributions of mock data contaminated with baryonic feedback falls within $0.3\sigma$ of the uncontaminated marginal posterior distribution (this criterion derived from the standard DES Y3 test).}
    \label{fig:baryons_test}
\end{figure}

We use 118 simulations from the \texttt{CosmoGridV1} simulation suite \citep{cosmogridKacprzak_2023} as an independent test set.
The simulations all share the same fiducial cosmology: $\sigma_8=0.84,\  \Om=0.26,\ w=-1,\ H_0 = 67.36~\mathrm{km\,s^{-1}\,Mpc^{-1}},\ \Omega_{\rm b} = 0.0493,\ n_s=0.9649$, and were created using the \pkdgrav{} code \citep{potter2017pkdgrav3}. To check for systematic errors in our analysis, we generate two sets of seed-matched mock data; one with a given systematic and one without. To marginalize over noise realizations, we compute posteriors on the \textit{average} statistics $\langle \textbf{t} \rangle_{i=1}^{N}$ from the contaminated and non-contaminated sets, following 
\citet{jeffrey2024dark}. To pass this test, the posterior of the averaged, contaminated statistic should not shift by more than $0.3\sigma$ in the $\Om$--$S_8$ plane and $w$ marginal, in line with the standard DES criterion.

\begin{figure*}
    \centering
    \includegraphics[width=0.85\linewidth]{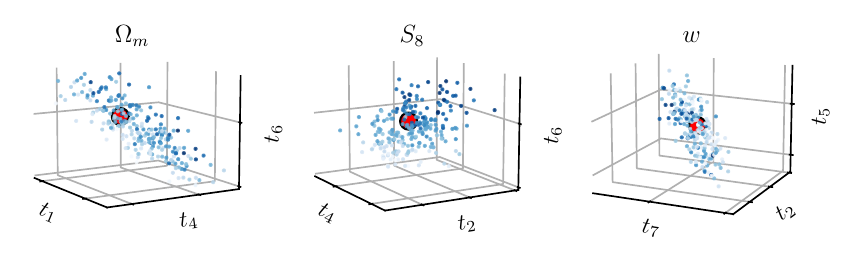}
    \caption{Three dimensional slices of hybrid summaries from simulations (scatter points) capture informative structure with respect to indicated parameter values (scatter colour gradient). The DES Y3 target data (red dot) falls squarely on this learned manifold, indicating that the compressed data is in-distribution with compressed simulations. This visual inspection is a blinded test, as it does not require assigning a cosmology to the target data.}
    \label{fig:summary_scatter}
\end{figure*} 

Baryonic feedback effects from galaxy formation can lead to the apparent suppression of cosmological structure, and are the dominant form of systematic uncertainty at small scales in weak lensing. The training simulations are dark-matter-only and do not include baryonic physics. The standard DES approach is to cut physical scales likely affected by baryons  and ensure that introducing a change in systematic prescription does not contaminate the cosmological results. Full forward baryonic feedback models require expensive and ill-specified hydrodynamical simulations  over the full cosmological volume, which are not available for DES-like volumes. The \texttt{CosmoGridV1}  simulations employ a baryon correction model, which changes the N-body density fields in post-processing to  emulate baryon feedback \citep{cosmogridKacprzak_2023}. We observe a very small effect of the baryonic feedback systematic on the posterior distribution of the averaged hybrid summaries (Fig.~\ref{fig:baryons_test}). This test was also passed for the  $C_\ell$-alone compression. 

\subsection{Posterior predictive distributions}
As an additional validation of our inference pipeline, we perform a `modular' posterior predictive distribution (modular PPD) test to assess whether the forward model adequately describes the observed DES Y3 data at the angular power spectra level. This test provides a direct check for model misspecification by comparing observables generated from the inferred posterior with the actual observations.

For the test, we generate samples of the noise-corrected lensing pseudo-$C_\ell$ power spectra from the posterior inferred from DES Y3 data. Rather than running new forward simulations, we leverage our existing test simulations by selecting those whose cosmological parameters are consistent with the inferred posterior. We sample from our posterior distribution $p(\parvec \cond \textbf{t}_{\rm DES})$ obtained from the DES Y3 hybrid summary statistics values $\textbf{t}_{\rm DES}$, giving a distribution of test simulations from our suite. We achieve this using the same rejection sampling method as described in Subsection~\ref{subsec:coverage}.

These simulations are still varied over the full prior $p(\boldsymbol{\eta})$ of nuisance parameters. The power spectra associated with these selected simulations $\dat$ are then draws from the \textit{modular} posterior predictive distribution:
\begin{equation}
    p(\dat \cond \textbf{t}_{\rm DES}) = \iintt p(\dat \cond \parvec, \boldsymbol{\eta}) p(\parvec \cond \textbf{t}_{\rm DES})p(\boldsymbol{\eta}) \, \mathd \parvec \, \mathd \boldsymbol{\eta}.
\end{equation}
We refer to this PPD as `modular' because in our SBI analysis we infer only $\Om$, $S_8$, and $w$, marginalizing over all other parameters (both nuisance and cosmological); all non-inferred quantities must therefore be varied over their respective priors when drawn for PPD comparison. For a further discussion see Appendix~\ref{app:mod_ppd}. 

This test is particularly informative because the uncompressed power spectra themselves are not directly used in our hybrid statistics inference; rather, they serve as an independent cross-check of the model at the level of two-point statistics. If the forward model accurately captures the data-generating process, the observed DES Y3 pseudo-$C_\ell$ measurements should fall within the posterior predictive distribution.

\begin{figure}
    \centering
    \includegraphics[width=\linewidth]{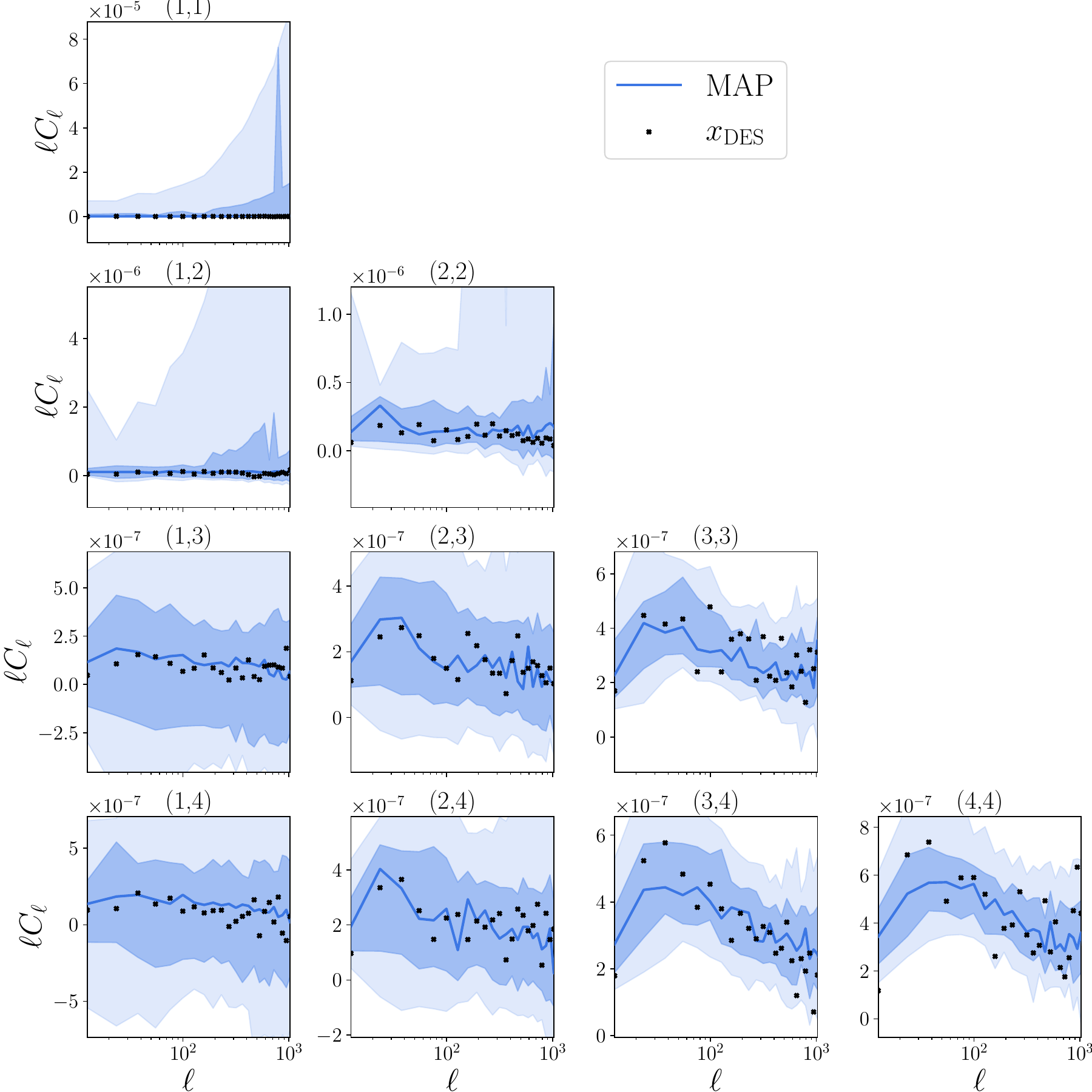}
    \caption{Modular posterior predictive distribution (PPD) test for the noise-subtracted DES Y3 pseudo-$C_\ell$ power spectra. The shaded regions show the $1\sigma$ and $2\sigma$ credible intervals of the posterior predictive distribution. The DES Y3 measurements fall comfortably within the predicted distributions across all tomographic bin combinations.}
    
    \label{fig:ppd}
\end{figure}

Figure~\ref{fig:ppd} shows the posterior predictive distributions of the pseudo-power spectra alongside the DES Y3 measurements. We compute the mean and the $1\sigma$ and $2\sigma$ credible intervals of the modular PPD $p(\dat \cond \textbf{t}_{\rm DES})$ in each $\ell$-bin. Across all tomographic bin combinations and angular scales included in our analysis, the observed power spectra are comfortably contained within the credible regions of the posterior predictive distributions. We conclude that there is no evidence that the Gower Street simulation suite, combined with our forward modelling, results in any model misspecification for the DES Y3 weak lensing data within the predictive power of our modular PPD test at the power spectrum level. In future studies, one could use higher-order statistics, such as peaks or higher-order moments, to do a PPD test beyond the power spectrum.

In Fig.~\ref{fig:ppd} the closer tomographic bins show very large credible intervals, particularly at smaller scales. This reflects the `modularity' of our PPD test as discussed above. Recall we are marginalizing over our uninferred cosmology and nuisance parameters; the relatively conservative choice of prior for NLA intrinsic alignment model parameters $A_{\textrm{IA}}$ and $\eta_{\textrm{IA}}$ results in a particularly large variance of power spectra at low-redshift tomographic bins and small scales. The farther tomographic bins are less affected by this choice of prior. Including these nuisance parameters in the inferred posteriors (rather than marginalizing over them) in a future analysis with a larger simulation budget will give tighter constraints on the PPD test.
 
\subsection{Blinding strategy}

In order to mitigate confirmation bias we follow the same blinding procedure as in \cite{jeffrey2024dark}. Whilst developing the analysis procedure, we remained blind to any cosmology results that could introduce a bias in the DES Y3 data results. We trained and validated the neural compression and amortized NDE pipeline on simulations alone before unblinding. All analysis choices relating to the impact of systematics testing and validating the pipeline were also conducted blindly on simulations. 

\begin{figure}
    \centering
    \includegraphics[width=0.85\linewidth]{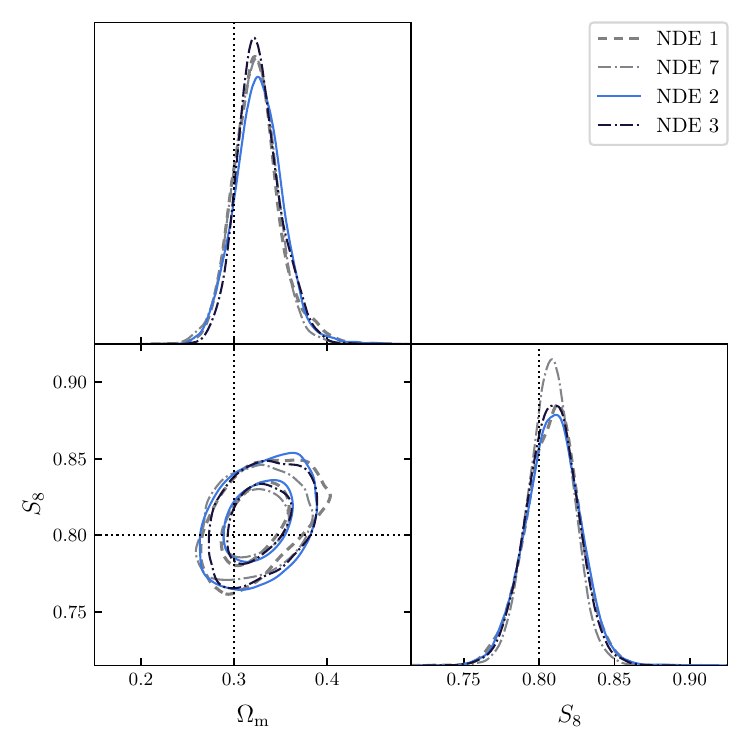}
    \caption{Neural density estimator convergence test on $\Om$ and $S_8$ for four (out of eight total) individual NDE ensemble members. This test was performed on the DES Y3 data by blindly shifting the mean value of the posteriors to fiducial values of $\Om=0.3$ and $S_8=0.8$. }
    \label{fig:nde_test}
\end{figure}

We performed two pre-unblinding checks:
\begin{itemize}[wide]
\item{
    We check that our DES Y3 summary statistics lie within the distribution of summary statistics from the Gower Street mocks, examples of which are displayed in Fig.~\ref{fig:summary_scatter}. In this plot, we colour simulations by their true cosmological parameters to demonstrate the learned structure in the summary statistics manifold. The target data (red dot) falls within the manifold, indicating that our data is qualitatively in-distribution. 
}
\item{
    We ensure that each of the individual NDEs of our ensemble converges when applied to the target data. To stay blind to the inferred posterior, output posterior chains for each ensemble member are mean-subtracted and shifted to the fiducial cosmology $(\Om, S_8)=(0.3,0.8)$, shown in Fig.~\ref{fig:nde_test}. To pass this test, each NDE posterior had to exhibit stable behaviour and good, mutual agreement with the rest of the ensemble.
}
\end{itemize}
Once these tests were passed we proceeded with unblinding.

\subsection{Post-unblinding changes}
\label{subsec:postunblindingchanges}

After unblinding it became apparent that the \TARP{} test results were not as well calibrated as had been thought (a software bug had led to this misapprehension). In reality, the \TARP{} test showed (some) evidence for both overconfidence and bias, primarily in the $\Om$--$w$ two-dimensional marginal posterior distribution. Investigation showed this to be due to the ensemble MAFs being poor at conditioning the likelihood on cosmology.

We implemented two changes post-unblinding:
\begin{itemize}[wide]
    \item The \VICreg{} regularization term was added to the final compression step (step~\ref{item:finalcompression} in Subsection~\ref{subsec:hierarchicalcompression}); pre-unblinding, we had used only the VMIM loss function. See Subsection~\ref{subsec:vicreg} for a discussion. This regularization did not affect the VMIM optimization objective; the final compression was found to converge to the same optimal loss as the pre-unblinding network. The regularization resulted in a likelihood that was easier both to estimate from the NDE base distributions and to condition on the parameters when training the NDEs. We found that compressing with only the variance and covariance regularization terms (with their weights set to $0.2$ and the invariance term discarded) produces summaries that are as invariant as those obtained when the invariance term is included in the optimization. However, we note that if the invariance term has a non-zero weighting, the learned summaries are more covariant than the pre-unblinding summaries. The measured variance and covariance of the post-unblinding summaries do not differ significantly from those of the pre-unblinding summaries. Therefore, we conclude that the invariance of the summaries, despite invariance not being an active term in the \VICreg{} regularization, allows the NDEs to learn the conditioning relationship more easily.
    
    \item{We increased the size of the ensemble of MAFs from 4 to 8, we increased the size of the conditioning networks inside the MAFs from 70 or 80 to 200, and we increased the number of MADE blocks from 7 or 8 to a more varied 6, 7, 8, or 9 transformations. These changes allow the network, when estimating the likelihood, to learn more complex conditional relationships between summary distributions and cosmology.}
\end{itemize}

The amendments resulted in only a small change to our cosmological constraints: all two-dimensional marginalized posteriors shifted by less than $0.3\sigma$, and there was marginally reduced precision in $\Om$ and $w$. See Appendix~\ref{app:analysis-change} (in particular Fig.~\ref{fig:analysis-change-posteriors}) for a comparison of the pre- and post-unblinding posterior distributions. In any case, we emphasize that the post-unblinding code changes were not motivated by misgivings about the inferred cosmological constraints.

\begin{figure}
    \centering
    \includegraphics[width=\linewidth]{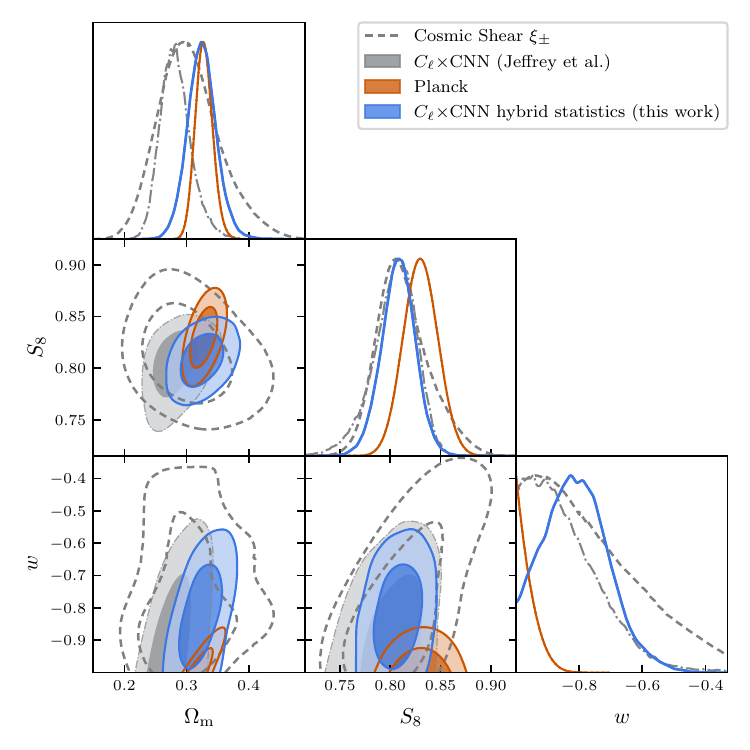}
    \caption{\textbf{Main result.} DES Y3 $w$CDM constraints: posterior probability distribution for parameters $\{ \Om, S_8, w \}$ with DES Y3 data. Hybrid statistics improves information extraction about all three parameters over two-point \citep{Doux_2022} and the existing CNN compression \citep{jeffrey2024dark}. These constraints are consistent with \citet{Planck2018}, the latter recalculated here with our analysis priors.}

    \label{fig:main_results}
\end{figure}

\begin{table}
    \centering
    \renewcommand{\arraystretch}{1.2}
    \resizebox{\columnwidth}{!}{
    \begin{tabular}{cccc}
        \toprule
		   & DES Y3 $\xi_\pm$& $C_\ell \times$ CNN  & \textbf{Hybrid statistics}  \\ 
          & likelihood$^*$ & \citep{jeffrey2024dark}  & \textbf{(this work)} \\ 
		\midrule
		$\Om$ & $0.303^{+0.040}_{-0.051}$ & $0.283^{+0.020}_{-0.027}$ & $\mathbf{0.325 \pm 0.024}$ \\ 
		$S_8$ & $0.813^{+0.020}_{-0.029}$ & $0.804^{+0.025}_{-0.017}$ & $\mathbf{0.808 \pm 0.017}$ \\ 
		$w$ & $< -0.707$ & $< -0.804$ & $\mathbf{< -0.766}$ \\ 
		\midrule
        MAP($w$) & $-0.940$ & $-0.953$ & $\mathbf{-0.826}$ \\
        \midrule
		$\FOM(\Om, w)$ & $138$ & $384$ & $\mathbf{470}$ \\
		$\FOM(S_8, w)$ & $391$ & $389$ & $\mathbf{592}$ \\
		$\FOM(\Om, S_8)$ & $901$ & $1{,}885$ & $\mathbf{2{,}614}$ \\
		$\FOM(\Om, S_8, w)$ & $10{,}344$ & $18{,}430$ & $\mathbf{29{,}444}$ \\
		\bottomrule
        & $^*$reanalysed & &  \\
    \end{tabular}}
\caption{Comparison with existing analyses: 68 per cent credible intervals from the marginal posterior probability distributions of $\Om$, $S_8$, $w$. The central number (where given) is the mean. We compare our hybrid statistics (map-level) compression to the existing $C_\ell \times$CNN map-level compression from \citet{jeffrey2024dark} and to the standard DES weak gravitational lensing (two-point correlation function) likelihood. The standard DES likelihood has been reanalysed to match the prior choices employed here, in the same way as the comparison presented in \citet{jeffrey2024dark}. Figure of Merit ($\FOM$) summaries are calculated using equation (\ref{eq:FoM}), while marginal distribution credible intervals are calculated using \texttt{GetDist} \citep{lewis2019getdist}.
}
\label{tab:resultscomp}
\end{table}

\section{DES Y3 results}\label{sec:results}

\subsection{Cosmological constraints}
We show in Fig.~\ref{fig:main_results}, in Table~\ref{tab:resultscomp}, and in Fig.~\ref{fig:omS8marg} our main results: posterior distributions of cosmological parameters derived using our hybrid statistics methodology. The results show the benefit of optimized data compression, as the cosmological constraints are considerably more precise in all planes of our inferred parameters $\{\Om, S_8, w \}$ compared to previous analyses. We quantify the improvements in terms of the Figure of Merit, defined for two cosmological parameters $\theta_1$ and $\theta_2$ to be 
\begin{equation}
    \label{eq:FoM}
    \FOM(\theta_1, \theta_2) = \frac{1}{\sqrt{\det(C_{\theta_1, \theta_2})}}
\end{equation}
where $C_{\theta_1, \theta_2}$ is the covariance of the marginalized posterior between $\theta_1$ and $\theta_2$ (there is a similar definition for three parameters). A larger $\FOM$ is better. Referring to Table~\ref{tab:resultscomp}:
\begin{itemize}[wide]
\item {
    For the weak lensing parameter combination $\{ \Om, S_8 \}$, we improve the $\FOM$ over the standard two-point analysis by a factor of 2.90, and by 39 per cent when compared with the previous map-level compression scheme in \cite{jeffrey2024dark}.
    }
\item{
    For the dark energy parameter combination $\{ \Om, w \}$ we quote improvement factors of 3.41 and 1.22 over two-point and existing map-level compression, respectively.
    }
\item{
    For the full three-parameter combination $\{ \Om, S_8, w \}$, we improve the $\FOM$ over the standard two-point analysis by a factor of 2.85, and by 60 per cent when compared with the existing CNN result.
    }
\end{itemize}

We can directly compare the constraints to those from DES Y3 weak lensing (for the latter, we use the analytical two-point correlation function likelihood, but coupled now with the same analysis prior as used in this study). We note $\Omega_{\rm b}$ was changed to match that of the DES $3\times2$pt analysis \citep{DES_3x2pt}, as our analysis prior comes from Planck (see Section~\ref{sec:simsdata}).

Our results are consistent with those from the DES Y3 two-point weak lensing analysis, but are significantly more constraining, and are also consistent with Planck in both $S_8$ and $\Om$.

Our analysis prior on $w$ has a lower bound at $-1$, but the marginalized posterior density is not close to zero near this boundary. We therefore express constraints on $w$ as the upper limit of the 68 per cent credible interval. The Maximum a Posteriori (MAP) value of $w$ is $w=-0.826$, but the $\Lambda$CDM value of $-1$ is comfortably accommodated by the data.

\subsection{Comparison with other statistics}

\begin{figure}
    \centering
    \includegraphics[width=\linewidth]{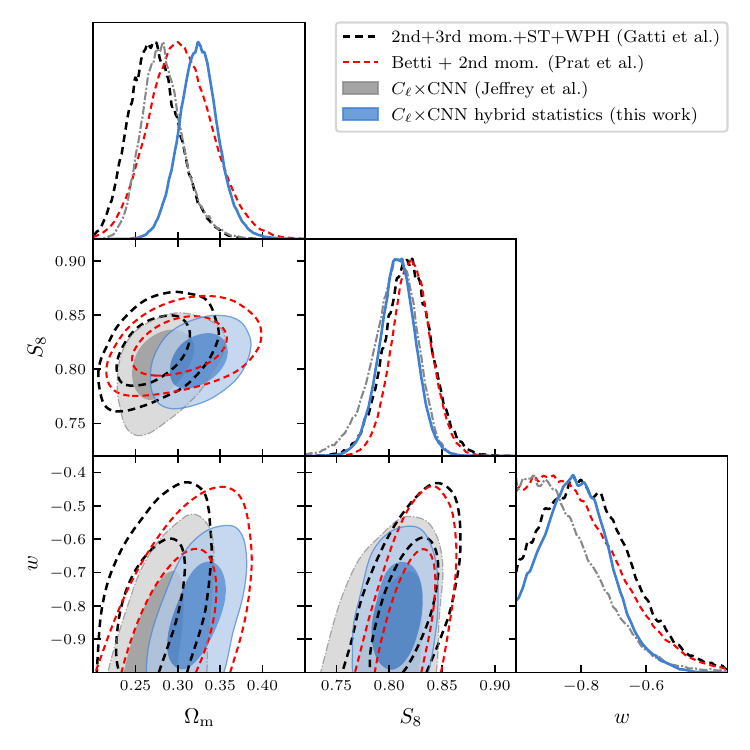}
    \caption{DES Y3 $w$CDM results: we compare the joint posterior probability distributions between this work's hybrid statistics and the results from \citet{gatti_dark_2024} which combines 2nd and 3rd order moments, scattering transforms and wavelet phase harmonics; \citet{juditprat_dark_2025} which combines Betti numbers and 2nd order moments; and \citet{jeffrey2024dark} which combines the angular power spectrum with a field-level CNN using a mean squared error as an optimization scheme.}
    \label{fig:sbi_compare_results}
\end{figure}

Figure~\ref{fig:sbi_compare_results} compares our results with those of three independent studies that use other higher-order statistics to analyse the same DES Y3 weak lensing data. All the studies (including ours) use the same simulations and same forward model for their SBI pipelines, including the same choice of analysis prior. Hence they are directly comparable, and their differences demonstrate the effect of changing the higher-order statistics used. 
\begin{itemize}[wide]
\item {
    \citet{gatti_dark_2024} use a combination of second- and third-order moments, scattering transform, and wavelet phase harmonics to achieve $\FOM(\Om,S_8)=1{,}725$.
}
\item {
    \citet{juditprat_dark_2025} use Betti numbers through persistent homology in combination with second order moments to achieve $\FOM(\Om,S_8)=1{,}542$.
}
\item {
    \citet{jeffrey2024dark} (included and summarized in Table~\ref{tab:resultscomp}; this study -- unlike the previous two  -- does not include the intrinsic alignment amplitude parameter in the inferred parameter data vector) perform field-level analysis to achieve $\FOM(\Om,S_8)=1{,}885$.
}
\item {
    The current study (Table~\ref{tab:resultscomp}) achieves $\FOM(\Om,S_8)=2{,}614$.
}
\end{itemize}

\section{Conclusions}\label{sec:conclusion}

We have presented a simulation-based inference analysis of the Dark Energy Survey Year 3 weak gravitational lensing data using a hybrid map-level compression scheme that combines angular power spectra with convolutional neural network (CNN) compression. Our approach uses information-theoretic arguments to optimize the extraction of cosmological information from weak lensing mass maps, seeking improvements in constraining power over existing analyses. The inclusion of map-level data substantially improves the constraints over the power spectra alone in both $\Lambda$CDM and $w$CDM.

Our simulation-based inference framework enables the forward modelling of realistic observational effects in the mock data, circumventing the need for an explicit likelihood function. Systematic effects that are challenging to model analytically, including sky masks, non-Gaussian shape noise, photometric redshift uncertainties, intrinsic galaxy alignments, and shear measurement biases, are propagated self-consistently in forward-modelled simulations. We have verified that our pipeline is robust to model misspecification in the form of baryon feedback contamination, that the resulting full (i.e. non-marginalized) posterior exhibits correct coverage, and that a posterior predictive test shows our model to be in good agreement with the data.

The hybrid statistics approach optimizes the neural compression of weak lensing maps to complement the cosmological parameter information already present in the angular power spectrum, rather than treating map-level and two-point information as independent. Using the same simulations and network architectures as \cite{jeffrey2024dark}, this `iterative compression scheme that builds upon the power spectrum information' yields substantially tighter constraints on the $w$CDM model parameters.  

Our analysis of the $w$CDM model with a prior $-1 < w < -1/3$ yields 68 per cent credible intervals of $\Om = 0.325^{+0.024}_{-0.024}$, $S_8 = 0.808^{+0.017}_{-0.017}$, and $w < -0.766$. These are the to-date most precise joint and marginal constraints of $\{ \Om, S_8 , w\}$ parameters from weak gravitational lensing alone. We quantify these constraints via the Figure of Merit ($\FOM$), finding for the three-parameter combination $\{\Om, S_8, w\}$ a nearly three-fold improvement over the standard DES Y3 two-point correlation function analysis and a factor of 1.6 improvement over the previous map-level CNN compression from \cite{jeffrey2024dark}, and, for the weak lensing parameter combination $\{\Om, S_8\}$, corresponding improvement factors of 3.0 and 1.4 over two-point statistics and existing map-level compression, respectively.  

Our results are the to-date most precise determination of cosmological parameters from weak lensing in the $w$CDM model (see Appendix~\ref{app:wl_comp} for a comparison with other weak lensing surveys). Our $\FOM(\Om, S_8) = 2{,}614$ is the highest of any weak lensing analysis. The results are consistent with $\Lambda$CDM and are in agreement with the Planck constraints (although, in common with other lensing analyses, the most probable value of $S_8$ is slightly low compared with Planck). They are also consistent with, but even more precise than, the latest results from the DES Year 6 $3\times2$pt analysis (\citealt{DESy6-3x2pt}; this latter analysis includes cosmic shear, galaxy–galaxy lensing, and galaxy clustering, and gives $S_8=0.782^{+0.021}_{-0.020}$, $\Om = 0.325^{+0.032}_{-0.035}$, $w=-1.12^{+0.26}_{-0.20}$); note, though, that the results cannot be directly compared as the latter analysis includes galaxy clustering probes and a slightly different choice of analysis priors.

The methodological advances presented in this work demonstrate the power of combining complementary summary statistics within a simulation-based inference framework. The success of our hybrid approach, which explicitly optimizes for information complementary to the power spectrum, suggests that further gains may be achievable through the combination of additional higher-order statistics such as scattering transforms and wavelets. This framework is readily applicable (e.g. to DES Y6, including time-dependent $w(z)$ as Gower Street 3.0 simulations become available) and scalable to forthcoming Stage IV weak lensing surveys from \textit{Euclid} and the Vera C.\ Rubin Observatory, where the increased statistical power will demand more sophisticated methods such as this for extracting cosmological information while maintaining robustness to systematic uncertainties. The principal challenge for such applications will be achieving sufficiently realistic data modelling in the forward simulations; if accomplished, this approach will increase the potential for cosmological discovery from weak gravitational lensing observations.

\section*{Data availability}
\begin{itemize}[wide]
\item{
See \url{www.star.ucl.ac.uk/GowerStreetSims/} for the Gower Street simulations.
}
\item{See \url{https://des.ncsa.illinois.edu} for the metacalibration lensing catalogue.
}
\item{The MCMC samples from the parameter posteriors (i.e. chains) will be made available upon publication of the accepted paper.
}
\end{itemize}

\section*{Contribution statement}

All authors contributed to this paper and/or carried out infrastructure work that made this analysis possible. Highlighted contributions include:
\begin{itemize}[wide]
    \item Algorithm design: T. L. Makinen, N. Jeffrey, A. Heavens, N. Porqueres, B. Wandelt
    \item Algorithm and analysis implementation: T.L. Makinen, J. Williamson
    \item Hybrid statistics project concept: T.L. Makinen, A. Heavens, N. Porqueres, B. Wandelt
    \item Simulation design and data pipeline: J. Williamson, M. Gatti, N. Jeffrey, T. L. Makinen, L. Whiteway
    \item Paper writing and figures: J. Williamson, T. L. Makinen, L. Whiteway, A. Heavens, N. Porqueres, N. Jeffrey, B. Wandelt, O. Lahav, M. Gatti, J. Prat
    \item Comments on manuscript: Z. Gong, A. Thomsen, C. Doux
    \item Final reader: A.R. Walker
\end{itemize}

\section*{Acknowledgements}

JW acknowledges the UK Science and Technology Facilities Council (STFC) grant ST/P006736/1. This work was supported by STFC through Imperial College Astrophysics Consolidated Grant ST/W000989/1. TLM acknowledges the Imperial College London President's Scholarship and InfoSys-Cambridge AI Centre for support of this work, as well as fruitful discussions with David Spergel and François Lanusse. NP is supported by the ERC StG grant 101220619 (OCAPi). NJ acknowledges support by the ERC-selected UKRI Frontier Research Grant EP/Y03015X/1.

The Gower Street simulations were generated under the DiRAC project dp153 `Likelihood-free inference with the Dark Energy Survey' (ACSP255/ACSC1) and dp327 (APP16265) using DiRAC (STFC) HPC facilities (\url{www.dirac.ac.uk}).

This research used resources of the National Energy Research Scientific Computing Center (NERSC), a U.S. Department of Energy Office of Science User Facility located at Lawrence Berkeley National Laboratory, operated under Contract No. DE-AC02-05CH11231 using NERSC award HEP-ERCAP-0027266.

Funding for the DES Projects has been provided by the U.S. Department of Energy, the U.S. National Science Foundation, the Ministry of Science and Education of Spain, 
the Science and Technology Facilities Council of the United Kingdom, the Higher Education Funding Council for England, the National Center for Supercomputing 
Applications at the University of Illinois at Urbana-Champaign, the Kavli Institute of Cosmological Physics at the University of Chicago, 
the Center for Cosmology and Astro-Particle Physics at the Ohio State University,
the Mitchell Institute for Fundamental Physics and Astronomy at Texas A\&M University, Financiadora de Estudos e Projetos, 
Funda{\c c}{\~a}o Carlos Chagas Filho de Amparo {\`a} Pesquisa do Estado do Rio de Janeiro, Conselho Nacional de Desenvolvimento Cient{\'i}fico e Tecnol{\'o}gico and 
the Minist{\'e}rio da Ci{\^e}ncia, Tecnologia e Inova{\c c}{\~a}o, the Deutsche Forschungsgemeinschaft and the Collaborating Institutions in the Dark Energy Survey. 

The Collaborating Institutions are Argonne National Laboratory, the University of California at Santa Cruz, the University of Cambridge, Centro de Investigaciones Energ{\'e}ticas, 
Medioambientales y Tecnol{\'o}gicas-Madrid, the University of Chicago, University College London, the DES-Brazil Consortium, the University of Edinburgh, 
the Eidgen{\"o}ssische Technische Hochschule (ETH) Z{\"u}rich, 
Fermi National Accelerator Laboratory, the University of Illinois at Urbana-Champaign, the Institut de Ci{\`e}ncies de l'Espai (IEEC/CSIC), 
the Institut de F{\'i}sica d'Altes Energies, Lawrence Berkeley National Laboratory, the Ludwig-Maximilians Universit{\"a}t M{\"u}nchen and the associated Excellence Cluster Universe, 
the University of Michigan, NSF NOIRLab, the University of Nottingham, The Ohio State University, the University of Pennsylvania, the University of Portsmouth, 
SLAC National Accelerator Laboratory, Stanford University, the University of Sussex, Texas A\&M University, and the OzDES Membership Consortium.

Based in part on observations at NSF Cerro Tololo Inter-American Observatory at NSF NOIRLab (NOIRLab Prop. ID 2012B-0001; PI: J. Frieman), which is managed by the Association of Universities for Research in Astronomy (AURA) under a cooperative agreement with the National Science Foundation.

The DES data management system is supported by the National Science Foundation under Grant Numbers AST-1138766 and AST-1536171.
Data access is enabled by Jetstream2 and OSN at Indiana University through allocation PHY240006: Dark Energy Survey from the Advanced Cyberinfrastructure Coordination Ecosystem: Services and Support (ACCESS) program, which is supported by U.S. National Science Foundation grants 2138259, 2138286, 2138307, 2137603, and 2138296.
The DES participants from Spanish institutions are partially supported by MICINN under grants PID2021-123012, PID2021-128989, PID2022-141079, SEV-2016-0588, CEX2020-001058-M and CEX2020-001007-S, some of which include ERDF funds from the European Union. IFAE is partially funded by the CERCA program of the Generalitat de Catalunya.

We acknowledge support from the Brazilian Instituto Nacional de Ci\^encia
e Tecnologia (INCT) do e-Universo (CNPq grant 465376/2014-2).

This document was prepared by the DES Collaboration using the resources of the Fermi National Accelerator Laboratory (Fermilab), a U.S. Department of Energy, Office of Science, Office of High Energy Physics HEP User Facility. Fermilab is managed by Fermi Forward Discovery Group, LLC, acting under Contract No. 89243024CSC000002.
Research leading to these results has received funding from the European Research
Council under the European Union's Seventh Framework Program (FP7/2007-2013) including ERC grant agreements 240672, 291329, and 306478.

This manuscript has been authored by Fermi Research Alliance, LLC under Contract No. DE-AC02-07CH11359 with the U.S. Department of Energy, Office of Science, Office of High Energy Physics.



\bibliographystyle{mnras_2author}
\bibliography{main,des_y3kp, bibliography}



\appendix

\section{Author affiliations} \label{append:affiliations}
{
\scriptsize
$^{1}$ Department of Physics \& Astronomy, University College London, Gower Street, London, WC1E 6BT, UK\\
$^{2}$ Imperial Centre for Inference and Cosmology (ICIC) \& Imperial Astrophysics, Imperial College London, Blackett Laboratory, Prince Consort Road, London SW7 2AZ, United Kingdom\\
$^{3}$ Department of Applied Mathematics and Theoretical Physics, University of Cambridge, Centre for Mathematical Sciences, Wilberforce Rd, Cambridge, CB3 0WA, UK\\
$^{4}$ Universit\'e Paris-Saclay, Universit\'e Paris Cit\'e, CEA, CNRS, AIM, 91191, Gif-sur-Yvette, France\\
$^{5}$ Department of Physics \& King's Institute for Artificial Intelligence, King's College London, Strand, London WC2R 2LS, United Kingdom\\
$^{6}$ Institute of Space Sciences (ICE, CSIC), Campus UAB, Carrer de Can Magrans, s/n, 08193 Barcelona, Spain\\
$^{7}$ Kavli Institute for Cosmological Physics, University of Chicago, Chicago, IL 60637, USA\\
$^{8}$ William H. Miller III Department of Physics \& Astronomy, Johns Hopkins University, 3400 N. Charles Street, Baltimore, MD 21218, USA\\
$^{9}$ Niels Bohr Institute, University of Copenhagen, Blegdamsvej 17, 2100 Copenhagen, Denmark\\
$^{10}$ Nordita, KTH Royal Institute of Technology and Stockholm University, Hannes Alfv\'ens v\"ag 12, SE-10691 Stockholm, Sweden\\
$^{11}$ University of Copenhagen, Dark Cosmology Centre, Juliane Maries Vej 30, 2100 Copenhagen \O, Denmark\\
$^{12}$ Department of Astrophysical Sciences, Princeton University, Peyton Hall, Princeton, NJ 08544, USA\\
$^{13}$ Physics Department, 2320 Chamberlin Hall, University of Wisconsin-Madison, 1150 University Avenue Madison, WI 53706-1390, USA\\
$^{14}$ Argonne National Laboratory, 9700 South Cass Avenue, Lemont, IL 60439, USA\\
$^{15}$ Department of Physics and Astronomy, University of Pennsylvania, Philadelphia, PA 19104, USA\\
$^{16}$ Department of Physics, Carnegie Mellon University, Pittsburgh, PA 15213, USA\\
$^{17}$ NSF AI Planning Institute for Physics of the Future, Carnegie Mellon University, Pittsburgh, PA 15213, USA\\
$^{18}$ Instituto de Astrofisica de Canarias, E-38205 La Laguna, Tenerife, Spain\\
$^{19}$ Laborat\'orio Interinstitucional de e-Astronomia - LIneA, Av. Pastor Martin Luther King Jr, 126 Del Castilho, Nova Am\'erica Offices, Torre 3000/sala 817, CEP: 20765-000, Rio de Janeiro, Brazil\\
$^{20}$ Universidad de La Laguna, Dpto. Astrof\'isica, E-38206 La Laguna, Tenerife, Spain\\
$^{21}$ Department of Physics, Duke University, Durham, NC 27708, USA\\
$^{22}$ NASA Goddard Space Flight Center, 8800 Greenbelt Rd, Greenbelt, MD 20771, USA\\
$^{23}$ Lawrence Berkeley National Laboratory, 1 Cyclotron Road, Berkeley, CA 94720, USA\\
$^{24}$ Universit\'e Grenoble Alpes, CNRS, LPSC-IN2P3, 38000 Grenoble, France\\
$^{25}$ Department of Astronomy and Astrophysics, University of Chicago, Chicago, IL 60637, USA\\
$^{26}$ Fermi National Accelerator Laboratory, P. O. Box 500, Batavia, IL 60510, USA\\
$^{27}$ California Institute of Technology, 1200 East California Blvd, MC 249-17, Pasadena, CA 91125, USA\\
$^{28}$ SLAC National Accelerator Laboratory, Menlo Park, CA 94025, USA\\
$^{29}$ University Observatory, Faculty of Physics, Ludwig-Maximilians-Universit\"at, Scheinerstr. 1, 81679 Munich, Germany\\
$^{30}$ Center for Astrophysical Surveys, National Center for Supercomputing Applications, 1205 West Clark St., Urbana, IL 61801, USA\\
$^{31}$ Department of Astronomy, University of Illinois at Urbana-Champaign, 1002 W. Green Street, Urbana, IL 61801, USA\\
$^{32}$ Department of Physics, ETH Zurich, Wolfgang-Pauli-Strasse 16, CH-8093 Zurich, Switzerland\\
$^{33}$ Kavli Institute for Particle Astrophysics \& Cosmology, P. O. Box 2450, Stanford University, Stanford, CA 94305, USA\\
$^{34}$ Instituto de F\'isica Gleb Wataghin, Universidade Estadual de Campinas, 13083-859, Campinas, SP, Brazil\\
$^{35}$ Department of Physics, University of Genova and INFN, Via Dodecaneso 33, 16146, Genova, Italy\\
$^{36}$ Jodrell Bank Centre for Astrophysics, School of Physics and Astronomy, University of Manchester, Oxford Road, Manchester, M13 9PL, UK\\
$^{37}$ Departament de F\'{\i}sica, Universitat Aut\`{o}noma de Barcelona (UAB), 08193 Bellaterra, Barcelona, Spain\\
$^{38}$ Institut de F\'{\i}sica d'Altes Energies (IFAE), The Barcelona Institute of Science and Technology, Campus UAB, 08193 Bellaterra (Barcelona), Spain\\
$^{39}$ Centro de Investigaciones Energ\'eticas, Medioambientales y Tecnol\'ogicas (CIEMAT), Madrid, Spain\\
$^{40}$ Brookhaven National Laboratory, Bldg 510, Upton, NY 11973, USA\\
$^{41}$ Department of Physics and Astronomy, Stony Brook University, Stony Brook, NY 11794, USA\\
$^{42}$ Institut de Recherche en Astrophysique et Plan\'etologie (IRAP), Universit\'e de Toulouse, CNRS, UPS, CNES, 14 Av. Edouard Belin, 31400 Toulouse, France\\
$^{43}$ Excellence Cluster Origins, Boltzmannstr.\ 2, 85748 Garching, Germany\\
$^{44}$ Max Planck Institute for Extraterrestrial Physics, Giessenbachstrasse, 85748 Garching, Germany\\
$^{45}$ Universit\"ats-Sternwarte, Fakult\"at f\"ur Physik, Ludwig-Maximilians Universit\"at M\"unchen, Scheinerstr. 1, 81679 M\"unchen, Germany\\
$^{46}$ Institute for Astronomy, University of Edinburgh, Edinburgh EH9 3HJ, UK\\
$^{47}$ Cerro Tololo Inter-American Observatory, NSF's National Optical-Infrared Astronomy Research Laboratory, Casilla 603, La Serena, Chile\\
$^{48}$ INAF-Osservatorio Astronomico di Trieste, via G. B. Tiepolo 11, I-34143 Trieste, Italy\\
$^{49}$ Physik-Institut, University of Z\"urich, Winterthurerstrasse 190, CH-8057 Z\"urich, Switzerland\\
$^{50}$ School of Mathematics and Physics, University of Queensland, Brisbane, QLD 4072, Australia\\
$^{51}$ Oxford College of Emory University, Oxford, GA 30054, USA\\
$^{52}$ Institut d'Estudis Espacials de Catalunya (IEEC), 08034 Barcelona, Spain\\
$^{53}$ Department of Physics, IIT Hyderabad, Kandi, Telangana 502285, India\\
$^{54}$ Instituto de Fisica Teorica UAM/CSIC, Universidad Autonoma de Madrid, 28049 Madrid, Spain\\
$^{55}$ Santa Cruz Institute for Particle Physics, Santa Cruz, CA 95064, USA\\
$^{56}$ Australian Astronomical Optics, Macquarie University, North Ryde, NSW 2113, Australia\\
$^{57}$ Lowell Observatory, 1400 Mars Hill Rd, Flagstaff, AZ 86001, USA\\
$^{58}$ George P. and Cynthia Woods Mitchell Institute for Fundamental Physics and Astronomy, and Department of Physics and Astronomy, Texas A\&M University, College Station, TX 77843, USA\\
$^{59}$ Aix Marseille Univ, CNRS/IN2P3, CPPM, Marseille, France\\
$^{60}$ Instituci\'o Catalana de Recerca i Estudis Avan\c{c}ats, E-08010 Barcelona, Spain\\
$^{61}$ Department of Physics, University of Cincinnati, Cincinnati, Ohio 45221, USA\\
$^{62}$ Ruhr University Bochum, Faculty of Physics and Astronomy, Astronomical Institute, German Centre for Cosmological Lensing, 44780 Bochum, Germany\\
$^{63}$ Computer Science and Mathematics Division, Oak Ridge National Laboratory, Oak Ridge, TN 37831, USA\\
$^{64}$ Berkeley Center for Cosmological Physics, Department of Physics, University of California, Berkeley, CA 94720, USA\\
}

\newpage

\section{Post-unblinding analysis comparison}\label{app:analysis-change}

\begin{figure}
    \centering
    \includegraphics[width=\linewidth]{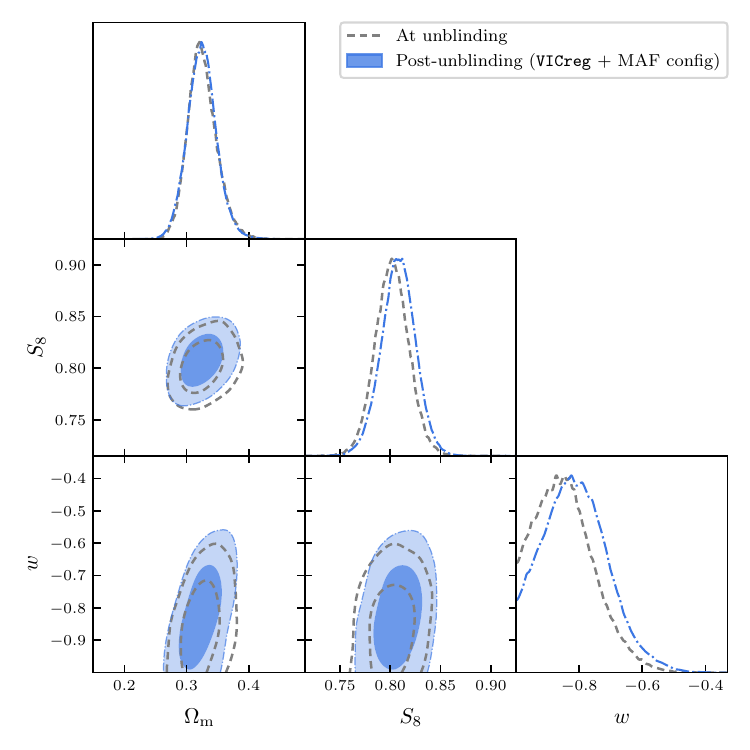}
    \caption{Comparison of cosmological constraints derived using pre- and post-unblinding code. See Subsection~\ref{subsec:postunblindingchanges} for details of the code changes.}
    \label{fig:analysis-change-posteriors}
\end{figure}

Subsection~\ref{subsec:postunblindingchanges} describes changes to the analysis pipeline made after unblinding. See both Fig.~\ref{fig:analysis-change-posteriors} and Table~\ref{tab:analysis-change-posteriors} for a comparison of the resulting pre- and post-unblinding cosmological constraints. Note that all three two-dimensional marginal distributions changed by at most $0.3\sigma$. 

\begin{table}
\caption {Changes to cosmological constraints arising from the post-unblinding code changes. See Subsection~\ref{subsec:postunblindingchanges} for details of the code changes.}
\centering
\begin{tabular}{ c c c c }
\toprule
\textbf{Parameter} & \textbf{Before} & \textbf{After} & \textbf{Abs Change}\\
\midrule
$\Om$   & $0.326^{+0.020}_{-0.026}$ & $0.325 \pm 0.024$ & $0.04\sigma$\\
$S_8$   & $0.802 \pm 0.017$ & $0.808 \pm 0.017$ & $0.35\sigma$\\
$w$          & $< -0.807$ & $< -0.766$\\
$\MAP(w)$    & $-0.846$ & $-0.826$\\
$\FOM(\Om,S_8,w)$ & $29{,}618$ & $29{,}444$ & $0.6$ per cent\\
$\Om$--$S_8$ & & & $0.28\sigma$\\
$\Om$--$w$ & & & $0.27\sigma$\\
$S_8$--$w$ & & & $0.30\sigma$\\
\bottomrule
\end{tabular}
\label{tab:analysis-change-posteriors}
\end{table}

\section{Modular posterior predictive}\label{app:mod_ppd}

Let $\dat^{\rm obs}$ denote observed data and $\dat$ a replicated future draw used for posterior predictive checks.

We partition the model parameters into two subsets $\parvec$ and $\boldsymbol{\eta}$, which are model parameters we choose to infer in the analysis and model parameters we marginalized over, respectively. The posterior distribution will be $p(\parvec, \boldsymbol{\eta} \cond \dat^{\rm obs})$, while the posterior predictive distribution (PPD) will be
\begin{equation}
p(\dat \cond \dat^{\rm obs}) = \iintt p(\dat \cond \parvec, \boldsymbol{\eta}) \, p(\parvec, \boldsymbol{\eta} \cond \dat^{\rm obs}) \, \mathd \parvec \, \mathd \boldsymbol{\eta}.
\end{equation}
This is the full Bayesian predictive distribution used in the usual goodness-of-fit checks. Discrepancies are evaluated by comparing $\dat^{\rm obs}$ to draws $\dat \sim p(\cdot \cond \dat^{\rm obs})$.

Now instead consider
\begin{equation}
p_{\textrm{mod}}(\dat \cond \dat^{\rm obs}) = \iintt p(\dat \cond \parvec, \boldsymbol{\eta}) \, p(\parvec \cond \dat^{\rm obs}) \, p(\boldsymbol{\eta}) \, \mathd \parvec \, \mathd \boldsymbol{\eta},
\end{equation}
which we can call the \textit{modular (cut) posterior predictive}. Here $\parvec$ is updated using $\dat^{\rm obs}$, but $\boldsymbol{\eta}$ is intentionally not updated (and is instead kept at its prior). Thus, $p_{\textrm{mod}}$ is a type of predictive distribution, but in general it differs from the full PPD $p(\dat \cond \dat^{\rm obs})$.

We apply this with $\parvec$ being the target parameters and $\boldsymbol{\eta}$ the nuisance parameters (i.e. the model parameters that we marginalize over). Even if SBI learns the marginal likelihood $p(\dat \cond \parvec) = \int p(\dat \cond \parvec, \boldsymbol{\eta}) p(\boldsymbol{\eta}) \, \mathd \boldsymbol{\eta}$ thereby yielding an unbiased posterior $p(\parvec \cond \dat^{\rm obs})$, we are generally unable to perform a full posterior predictive check for the original $(\parvec, \boldsymbol{\eta})$ model, since we do not obtain the updated nuisance parameter distribution $p(\boldsymbol{\eta} \cond \parvec, \dat^{\rm obs})$.

\section{Comparison to other lensing analyses}\label{app:wl_comp}

\begin{figure*}
    \centering
    \includegraphics[width=\textwidth]{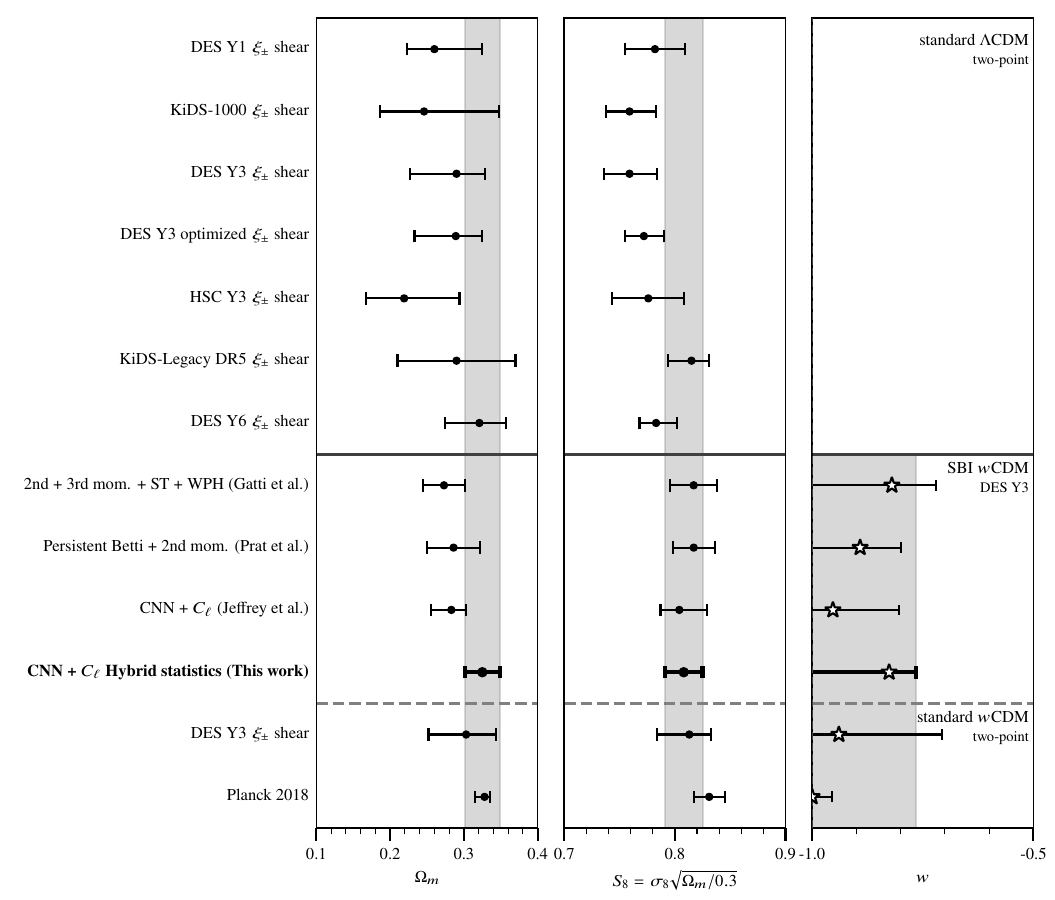}
    \caption{Marginal one sigma constraints for weak lensing parameters $\Om$, $S_8$, and $w$ from cosmic shear alone and including Planck 2018 TTTEEE+lowE likelihood for comparison. Hybrid statistics yield the to-date tightest marginal constraints from weak lensing data alone. Constraints for the $w$CDM model are with priors equivalent to those used in this analysis (apart from $\Omega_{\rm b}$ in Planck 2018) and are directly comparable. The results in the $\Lambda$CDM portion are listed for qualitative comparison, as the model and prior choices differ in these analyses. The means of the $\Om$ and $S_8$ marginal posterior distributions are indicated with a solid black circle. In common with the other lensing analyses, $S_8$ is low compared with Planck, but not significantly so. The $w$ marginal posterior distribution constraints are expressed as an upper bound (as our prior truncates at a point of non-negligible posterior density). The maximum a posteriori value of $w$ is marked with a star. The error bars show the mean and $68$ per cent credible interval of the marginal posterior distribution. These results are from \citet{ desY1shear, 
    desY3shear, desY3optimal, hscY3, Wright_2025, St_lzner_2025, DES2026shear}.
    }     
    \label{fig:omS8marg}
\end{figure*}

Figure~\ref{fig:omS8marg} compares posterior marginal one sigma credible regions for $\Om$, $S_8$, and $w$: from this study, from existing DES Y3 studies using weak lensing higher-order statistics, from the Planck TTTEEE+lowE and DES Y3 shear two-point data (reanalysed individually with priors matching those of this analysis), and from various existing analyses of $\Lambda$CDM (included for qualitative comparison).



\bsp	
\label{lastpage}
\end{document}